\documentclass[preprint,aps,prc,showpacs,showkeys]{revtex4}

\usepackage{graphicx}
\usepackage{bm}

\begin{document}

\title{ The $N N \to N \Delta$ cross section in nuclear matter%
        \footnote{Supported by BMBF and GSI Darmstadt} }

\author{ A.B. Larionov%
         \footnote{On leave from RRC "I.V. Kurchatov Institute",
                   123182 Moscow, Russia}
         and U. Mosel }

\affiliation{ Institut f\"ur Theoretische Physik, Universit\"at Giessen,
          D-35392 Giessen, Germany }

\date{\today}

\begin{abstract}
We present calculations of the $N N \to N \Delta$ cross section 
in nuclear matter within the one-pion exchange model taking into 
account pion collectivity, vertex renormalization by the contact nuclear
interactions and Dirac effective masses of the baryons due to coupling 
with the scalar $\sigma$ field. Introducing the Dirac effective masses 
leads to an in-medium reduction of the cross section. The experimental 
data on pion multiplicities from the collisions of Ca+Ca, Ru+Ru and Au+Au at 
$0.4\div1.5$ A GeV are well described by BUU calculations with 
the in-medium cross section.
\end{abstract}



\maketitle

\section{ Introduction }

While transport calculations have, in general, been very successful
in describing particle production in heavy-ion collisions 
\cite{BDG88,Cas90,Mosel91,Ehe93,Bass95,Ko96,Teis97,HR97,Uma98,CB99,Weber02},
one of the remaining open problems in the transport-theoretical description 
of heavy-ion collisions at SIS energies ($\sim 1$ A GeV) is an overprediction
of the pion multiplicity \cite{Bass95,Teis97,HR97,Uma98,Pelte97,LCLM01}. 
At the beam energies of a few A GeV the dominating mechanism of the pion 
production is the
excitation of the $\Delta(1232)$ resonance in a nucleon-nucleon (NN)
collision followed by its decay: $N N \to N \Delta$, $\Delta \to N \pi$.
The pion multiplicity, therefore, depends crucially on the value of
the in-medium $N N \to N \Delta$ cross section which -- by detailed
balance -- also determines the pion reabsorption. In Ref. \cite{LCLM01}
a phenomenological density-dependent modification factor of the vacuum
$N N \to N \Delta$ cross section has been proposed. The BUU calculations
with the density dependent $N N \leftrightarrow N \Delta$ cross sections 
are in a reasonable agreement with the experimental data on pion production
\cite{LCLM01}. 

The purpose of the present work is to get rid of the ad-hoc description
of Ref. \cite{LCLM01} and to calculate the in-medium 
$N N \leftrightarrow N \Delta$ cross sections on the basis of
the  one-pion exchange model (OPEM) with medium modifications.
The calculated cross sections are used then in the BUU model to 
study the pion production in heavy-ion collisions at SIS energies.

It has been shown in Ref. \cite{tHM87} within the
relativistic Dirac-Brueckner approach that the $N N \to N \Delta$
cross section is reduced at high densities. Later studies within the
OPEM including pion collectivity and vertex corrections 
\cite{Bertsch88,WuKo89,HR95} have shown, however, that the $N N \to N 
\Delta$ cross section grows with nuclear density. An important contribution 
which was missed in Refs. \cite{Bertsch88,WuKo89,HR95} is given by the Dirac
effective nucleon and $\Delta$ resonance masses. We will show, that the 
introduction of the Dirac effective nucleon and  $\Delta$ masses in the 
OPEM leads to an in-medium reduction of the  $N N \leftrightarrow N \Delta$ 
cross sections in agreement with Ref. \cite{tHM87}.

The structure of the paper is as follows: In Sect. II we describe the
OPEM with medium modifications including pion collectivity, vertex 
renormalization by the contact interactions and effective masses for 
the nucleon and $\Delta$ resonance. The energy dependence of the total 
$p p \to n \Delta^{++}$ cross section as well as the angular and 
$\Delta$ resonance mass dependencies of the differential 
$p p \to n \Delta^{++}$ cross section are studied at finite nuclear matter 
densities. Sect. III explains how the in-medium modified cross sections are 
implemented into the BUU code and describes the results of the BUU 
calculations of pion production from heavy-ion collisions at 0.5$\div$2 
A GeV. In sect. IV the summary of our results and of
remaining problems is given.

\section{ The model }

For the calculation of the $N N \to N \Delta$ cross section in nuclear
matter we apply the nonrelativistic OPEM similar to Ref. \cite{Bertsch88},
however, with readjusted parameters. We have chosen the nonrelativistic
version since it incorporates the contact nuclear interactions in a 
natural way, which is necessary in the in-medium calculations. 

In the nonrelativistic reduction (c.f. Ref. \cite{EW88}) the $\pi N N$ and
$\pi N \Delta$ interaction Lagrangians are:
\begin{eqnarray}
      {\cal L}_{\pi N N} & = & {f \over m_\pi} 
\psi^\dag \sigma_\alpha \mbox{\boldmath ${\mathbf \tau}$ \unboldmath} \psi 
\nabla_\alpha 
\mbox{\boldmath ${\mathbf \pi}$ \unboldmath}~,        \label{Lpinn} \\
      {\cal L}_{\pi N \Delta} & = & {f_\Delta \over m_\pi}
\psi^\dag_\Delta S_\alpha {\bf T} \psi 
\nabla_\alpha
\mbox{\boldmath ${\mathbf \pi}$ \unboldmath} + h.c.~,  \label{Lpind}
\end{eqnarray}
where $\psi,~\psi_\Delta$ and \boldmath ${\mathbf \pi}$ \unboldmath
are the nucleon, $\Delta$ resonance and pion field respectively.
$f = 1.008$ and $f_\Delta = 2.202$ are the coupling constants 
\cite{Dm86}. $\sigma_\alpha$ and 
\boldmath ${\mathbf \tau}$ \unboldmath are the spin and isospin Pauli
matrices. $S_\alpha$ and ${\bf T}$ are the spin and isospin transition
($1/2 \to 3/2$) operators defined according to Ref. \cite{BW75}.

The nuclear spin-isospin short range correlations are known to be
important for the in-medium $N N \to N \Delta$ cross section 
\cite{Bertsch88}. We will introduce them via the following Lagrangian:
\begin{eqnarray}
      {\cal L}_{SRC} &=& -{f^2 \over 2 m_\pi^2} g_{NN}^\prime
(\psi^\dag \sigma_\alpha \mbox{\boldmath ${\mathbf \tau}$ \unboldmath} \psi)
(\psi^\dag \sigma_\alpha \mbox{\boldmath ${\mathbf \tau}$ \unboldmath} \psi)
                                                 \nonumber \\
& & - \left[ {f f_\Delta \over m_\pi^2} g_{N\Delta}^\prime
(\psi^\dag \sigma_\alpha \mbox{\boldmath ${\mathbf \tau}$ \unboldmath} \psi)
(\psi^\dag_\Delta S_\alpha {\bf T} \psi) + \mbox{h.c.} \right] 
                                                 \nonumber \\
& & - {f_\Delta^2 \over m_\pi^2} g_{\Delta\Delta}^\prime
      (\psi^\dag_\Delta S_\alpha {\bf T} \psi)
      (\psi^\dag S_\alpha^\dag {\bf T}^\dag \psi_\Delta)
                                                 \nonumber \\
& & - \left[ {f_\Delta^2 \over 2 m_\pi^2} g_{\Delta\Delta}^\prime
      (\psi^\dag_\Delta S_\alpha {\bf T} \psi)
      (\psi^\dag_\Delta S_\alpha {\bf T} \psi) + \mbox{h.c.} \right]
                                                 \label{Lsrc}
\end{eqnarray}
$g_{NN}^\prime$, $g_{N\Delta}^\prime$ and $g_{\Delta\Delta}^\prime$ are
the Landau-Migdal parameters. 

The values of the Landau-Migdal parameters are not fixed unambiguously in 
the literature. Within a simple universality assumption 
$g_{NN}^\prime = g_{N\Delta}^\prime = g_{\Delta\Delta}^\prime
\equiv g_{BW}^\prime$, which is the B\"ackmann-Weise choice 
(c.f. Ref. \cite{MeyerTV81} and Refs. therein), one gets 
$g_{BW}^\prime = 0.7 \pm 0.1$ from the best description of the unnatural
parity isovector states in $^4$He, $^{16}$O and $^{40}$Ca. 
However, the same calculations within the Migdal model \cite{Mig90} 
assumption $g_{N\Delta}^\prime = g_{\Delta\Delta}^\prime = 0$ produce
$g_{NN}^\prime = 0.9\div1$. The description of the quenching of the
Gamow-Teller matrix elements requires 
$g_{\Delta\Delta}^\prime = 0.6\div0.7$ 
(assuming $g_{N\Delta}^\prime = g_{\Delta\Delta}^\prime$) \cite{Oset79}.
From the real part of the pion optical potential in $\pi$-atoms  one
gets $g_{N\Delta}^\prime = 0.2$ and $g_{\Delta\Delta}^\prime = 0.8$
\cite{Mig90}. The pion induced two-proton emission is the best described 
with $g_{N\Delta}^\prime = 0.25\div0.35$ \cite{Koerfgen97}.
We will adopt the two sets of the Landau-Migdal parameters from 
Ref. \cite{Arve94} (see Table 1).

As a first step, we calculate the $N N \to N \Delta$ cross section
in vacuum. The in-medium corrections are introduced in a second step.

\subsection{ Vacuum cross section }

The differential $N_1 N_2 \to N_3 \Delta_4$ cross section can be written
as:
\begin{equation}
        d\sigma = (2\pi)^4 \delta^{(4)}(p_1+p_2-p_3-p_4)
        \overline{|T|^2} { (2m_N)^3 2M_\Delta \over 4I }
        { d^3p_3 \over (2\pi)^3 2\varepsilon_3 }  
        { d^3p_4 \over (2\pi)^3 2\varepsilon_4 }
        {\cal A}_\Delta(M_\Delta^2) dM_\Delta^2~,           \label{sigvac}
\end{equation}
where $\overline{|T|^2}$ is the square of the absolute value of the reaction 
amplitude $T$ in the normalization of Ref. \cite{BD} averaged  
over spins of incoming particles 1,2 and summed over spins of outgoing 
particles 3,4. The flux factor is~: $I \equiv \sqrt{ (p_1p_2)^2 - m_N^4 }$.
The spectral function of the $\Delta$ resonance 
\begin{equation}
        {\cal A}_\Delta(M_\Delta^2) = {1 \over \pi}
        { M_\Delta \Gamma_\Delta(M_\Delta) \over 
          (M_\Delta^2-m_\Delta^2)^2 + 
           M_\Delta^2 \Gamma_\Delta^2(M_\Delta) }~,
                                                            \label{spfun}
\end{equation}
where $m_\Delta = 1.232$ GeV is the $\Delta$ pole mass, satisfies the 
normalization condition~:
\begin{equation}
       \int\limits_{(m_N+m_\pi)^2}^\infty\,dM_\Delta^2 
        {\cal A}_\Delta(M_\Delta^2) = 1~.
                                                            \label{Spnorma}
\end{equation}
The mass dependent total $\Delta$ width $\Gamma_\Delta(M_\Delta)$ 
is parameterized according to Ref. \cite{EBM99}~:
\begin{equation}
         \Gamma_\Delta(M_\Delta) 
       = \Gamma_\Delta^0\left({q(M_\Delta,m_N,m_\pi) 
                         \over q(m_\Delta,m_N,m_\pi)}\right)^3
         {m_\Delta \over M_\Delta} 
         {\beta_0^2+q^2(m_\Delta,m_N,m_\pi) 
    \over \beta_0^2+q^2(M_\Delta,m_N,m_\pi)}~, 
                                                             \label{gamdel}
\end{equation}
where $\Gamma_\Delta^0=0.118$ GeV is the width at the pole mass,
$\beta_0 = 0.2$ GeV/c is the cut-off parameter. In Eq.(\ref{gamdel}) 
and below
\begin{equation}
       q(m,m_1,m_2) = \sqrt{{(m^2+m_1^2-m_2^2)^2 \over 4m^2} - m_1^2}
                                                         \label{q}
\end{equation}
is the center-of-mass (c.m.) momentum of outgoing particles 
with masses $m_1$ and $m_2$ from the decay of a particle with mass $m$.  
In the c.m. frame of colliding nucleons one gets after standard 
transformations~:
\begin{equation}
        {d\sigma \over dM_\Delta^2 d\Omega} 
   = { (2m_N)^3 2M_\Delta \over 64 \pi^2 s} \overline{|T|^2}
     { q(\sqrt{s},M_\Delta,m_N) 
       \over q(\sqrt{s},m_N,m_N)} 
     {\cal A}_\Delta(M_\Delta^2)~,
                                                         \label{sigvaccm}
\end{equation}
where $s=(p_1+p_2)^2$.

\begin{figure}

\vspace{5cm}

\includegraphics{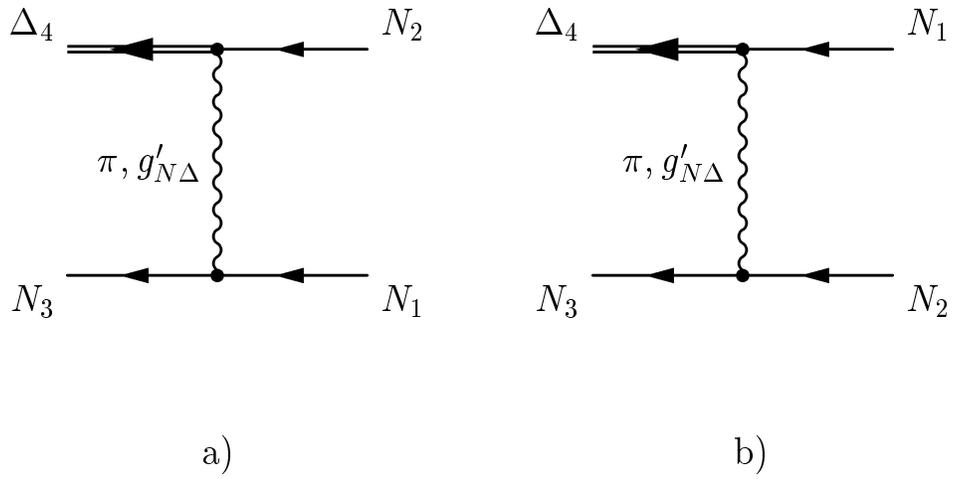}

\vspace{7cm}

\caption{\label{fig:nnnd_opem} Direct (a) and exchange (b) diagrams
contributing to the amplitude of the process $N_1 N_2 \to N_3 \Delta_4$.
The wiggly line denotes either $\pi$ exchange or the contact interaction
$\propto g_{N\Delta}^\prime$.} 
\end{figure}

The amplitude $T$ is evaluated from the graphs shown in 
Fig.~\ref{fig:nnnd_opem}.
In order to preserve the symmetry of the cross section with 
respect to $\Theta_{c.m.} = 90^o$ we have included the contact interaction
both in the direct and in the exchange diagrams. The same was done
in Refs. \cite{Bertsch88,Fernandez95} whereas in Ref. \cite{Koerfgen97} 
the contact interaction has been included in the direct term only. 
For the discussion on this subject we refer the
reader to the Appendix of Ref. \cite{Fernandez95}. Applying  Feynman
rules with the Lagrangians (\ref{Lpinn})-(\ref{Lsrc}) one gets the
following expressions for the direct (a) and exchange (b) terms of
the amplitude~:
\begin{eqnarray}
         iT_a & = & [\chi^\dag_\Delta(4) \Gamma_j^{abs}(k)  \chi(2)]  
                  iD^{(0)}(k) 
[\chi^\dag(3) \Gamma_j^{dec}(k) \chi(1)]  
     - iV_{4,3;2,1}(k)~,                                     \label{Ta} \\
         iT_b & = & -[\chi^\dag_\Delta(4) \Gamma_j^{abs}(k^\prime) \chi(1)]
                  iD^{(0)}(k^\prime) 
                    [\chi^\dag(3) \Gamma_j^{dec}(k^\prime)  \chi(2)]
     + iV_{4,3;1,2}(k^\prime)~.                              \label{Tb}
\end{eqnarray}
Here $\chi$ and $\chi_\Delta$ are the spin-isospin wave functions of
the nucleon and delta~:
\begin{equation}
   \chi = \delta_{m_t m} \delta_{\lambda s}~,~~~
   \chi_\Delta = \delta_{M_t M} \delta_{\lambda_\Delta s_\Delta}~,
                                                          \label{chis}
\end{equation}                 
where $m_t = \pm 1/2$ ($M_t = \pm 3/2,~\pm 1/2$) and $\lambda = \pm 1/2$
($\lambda_\Delta = \pm 3/2,~\pm 1/2$) are the quantum numbers of the
third isospin and spin component of a nucleon (delta). The quantities
$m$, $s$, $M$ and $s_\Delta$ are the corresponding indices of the wave 
functions. The four-momenta of the exchange pion in the direct and exchange 
channels are $k = p_1 - p_3$ and $k^\prime = p_2 - p_3$, respectively.
Below we use the kinematical invariants $t = k^2$ and $u = k^{\prime 2}$.
The operators 
\begin{eqnarray}
        \Gamma_j^{abs}(k) &=& -{f_\Delta(t) \over m_\pi}
  ({\bf S \cdot k}) T_j~,                       \label{gama0} \\
        \Gamma_j^{dec}(k) &=& {f(t) \over m_\pi}
  (\mbox{\boldmath${\mathbf\sigma}$\unboldmath}{\bf \cdot k}) \tau_j
                                                              \label{gamd0}
\end{eqnarray}
represent the $\pi N \to \Delta$ absorption and $N \to \pi N$ decay vertices
in vacuum. The quantity  
\begin{equation}
       V_{4,3;2,1}(k) = {f_\Delta(t)f(t) \over m_\pi^2} g_{N\Delta}^\prime
        [\chi^\dag_\Delta(4) S_\alpha {\bf T} \chi(2)]
        [\chi^\dag(3) \sigma_\alpha 
         \mbox{\boldmath${\mathbf\tau}$\unboldmath} \chi(1)]      
                                                             \label{Vcont0}
\end{equation}
is the contact interaction in vacuum and
\begin{equation}
           D^{(0)}(k) = {1 \over k^2 - m_\pi^2 + i0}     \label{piprop0}
\end{equation}
is the free pion propagator. In the $\pi N N$ and $\pi N \Delta$ vertices
we use the monopole form factor $F$~:
\begin{equation}
       f(t) = f F(t)~,~~~f_\Delta(t) = f_\Delta F(t)~,~~~
       F(t) = {\Lambda^2 - m_\pi^2 \over \Lambda^2 - t}    \label{ffact}
\end{equation}
with $\Lambda$ being the cut-off parameter (see Table 1). Following 
Refs. \cite{BW75,Bertsch88} a factor $F^2$ is included in the contact 
interaction vertices, since these vertices simulate the exchange by some 
heavy meson. 

The matrix element squared is
\begin{equation}
   |T|^2 = |T_a + T_b|^2 = |T_a|^2 + |T_b|^2 + T_a T_b^* + T_a^* T_b~.
                                                           \label{Tsq}
\end{equation}
We now consider the case of the $p p \to n \Delta^{++}$ reaction.
The spin-averaged direct term in (\ref{Tsq}) is
\begin{equation}
   {1 \over 4} \sum_{\lambda_1,\lambda_2,\lambda_3,\lambda_{\Delta4}}\,
   |T_a|^2  =  {4 \over 3} 
                 \left( {f_\Delta(t) f(t) \over m_\pi^2} \right)^2
                [ {\bf k}^4 D^{(0)2}(k)                     
                  +  2 {\bf k}^2 D^{(0)}(k) g_{N\Delta}^\prime 
                   + 3 (g_{N\Delta}^\prime)^2 ]~.        \label{Tasq}
\end{equation}
The spin-averaged exchange term 
${1 \over 4} \sum_{\lambda_1,\lambda_2,\lambda_3,\lambda_{\Delta4}}\,
|T_b|^2$ is given by Eq. (\ref{Tasq}) with replacements $k \to k^\prime$
and $t \to u$. The spin-averaged interference term can be written as
follows~:
\begin{eqnarray}
 &{1 \over 4}& \sum_{\lambda_1,\lambda_2,\lambda_3,\lambda_{\Delta4}}\,
[T_a T_b^* + T_a^* T_b] = - {f_\Delta(t)f(t)f_\Delta(u)f(u) \over 2m_\pi^4}
\left[ {4 \over 3}(({\bf k \cdot k^\prime})^2 + {\bf k}^2 {\bf k^\prime}^2)
  D^{(0)}(k)  D^{(0)}(k^\prime) \right.                    \nonumber \\
 &+& \left. {16 \over 3} {\bf k}^2 D^{(0)}(k) g_{N\Delta}^\prime 
  + {16 \over 3} {\bf k^\prime}^2 D^{(0)}(k^\prime) g_{N\Delta}^\prime
  + 16(g_{N\Delta}^\prime)^2 \right]~.                      \label{Tint}
\end{eqnarray}   

In numerical calculations we also took into account the c.m. 
and relativistic corrections to the amplitudes (\ref{Ta}),(\ref{Tb}) 
(c.f. Ref. \cite{Fernandez95})~:
(i) We replaced in (\ref{gama0}) $({\bf S \cdot k}) \to ({\bf S \cdot k_4})$ 
and $({\bf S \cdot k^\prime}) \to ({\bf S \cdot k^\prime_4})$ 
in the direct and exchange terms respectively. Here
${\bf k_4}$ (${\bf k_4^\prime}$) is the momentum of the exchanged pion
in the rest frame of $\Delta$ for the direct (exchange) term~:
\begin{eqnarray}
    {\bf k_4} &=& {\bf k} + \left(\left({E_4 \over M_\Delta} - 1\right)
   {{\bf k \cdot p_4} \over {\bf p_4}^2} - 
    {k_0 \over M_\Delta}\right){\bf p_4}~,                  \label{k4} \\
    {\bf k_4^\prime} &=& {\bf k^\prime} + 
    \left(\left( {E_4 \over M_\Delta} - 1\right)
   {{\bf k^\prime \cdot p_4} \over {\bf p_4}^2} - 
    {k_0^\prime \over M_\Delta}\right){\bf p_4}~.           \label{k4p}
\end{eqnarray}
(ii) The direct amplitude $T_a$ is multiplied by the factor 
$(-t/{\bf k}^2)^{1/2}$ and the exchange amplitude $T_b$
by the factor $(-u/{\bf k^\prime}^2)^{1/2}$. This restores approximately
the Lorentz invariance after summation of the amplitude squared over
spin projections.

After the c.m. and relativistic corrections Eqs.(\ref{Tasq}),(\ref{Tint})
get slightly modified. We do not show here explicit formulas: they will be 
given below for the more general in-medium case 
(see Eqs.(\ref{Tamedsq}),(\ref{Tintmedsq}) in Appendix A).

In the upper panel of Fig.~\ref{fig:sigma_vs_plab} we present the 
total cross section $\sigma_{pp \to n\Delta^{++}}$ as a function of the
c.m. energy. Dashed and solid lines correspond to the case of Set 1 
and Set 2, respectively; the dotted line represents the relativistic 
calculation using the model of Ref. \cite{Dm86}.
The relativistic and our corrected nonrelativistic calculations are in a 
good agreement at $\sqrt{s} < 3$ GeV. At higher c.m. energies 
our calculations produce somewhat larger cross section than the relativistic
result of Ref. \cite{Dm86}. In the lower panel of 
Fig.~\ref{fig:sigma_vs_plab} we show the total cross section of the 
process $\sigma_{pp \to pn\pi^+}$ vs the c.m. energy in comparison to the 
data from Ref. \cite{Bald87}. In this case the dominating channel is still 
$\sigma_{pp \to n\Delta^{++}},~\Delta^{++} \to p\pi^+$, however, also other 
channels contribute: $\sigma_{pp \to p\Delta^{+}},~\Delta^{+} \to n\pi^+$, 
which has a cross section of $1/9$ of the one for the channel through the 
$\Delta^{++}$ production; channels with higher resonance excitation 
\cite{Teis97}, whose 
contributions are shown by long-dashed and dash-dotted lines for the 
isospin 1/2 and 3/2 resonances, respectively; the nonresonant (background) 
$s$-wave channel $pp \to np\pi^+$ \cite{Teis97} shown by short-dashed line.
The sum of all contributions in the case of our $pp \to n\Delta^{++}$
cross section calculated with the Set 2 and in the case of the 
cross section from Ref. \cite{Dm86} is shown by the solid and dotted 
lines, respectively. The experimental data are well described in both cases.
For $\sqrt{s} < 3$ GeV the Set 2 is a little closer to the data, while 
for higher c.m. energies the relativistic calculation of Ref. \cite{Dm86}
works somewhat better. For the practical applications the difference at 
high c.m. energies is not important, since anyway the FRITIOF model 
is applied for the baryon-baryon collisions 
at $\sqrt{s} > 2.6$ GeV in the BUU program \cite{EBM99} which we use
in the description of pion production from heavy-ion collisions.  

\begin{figure}

\vspace{1cm}

\includegraphics{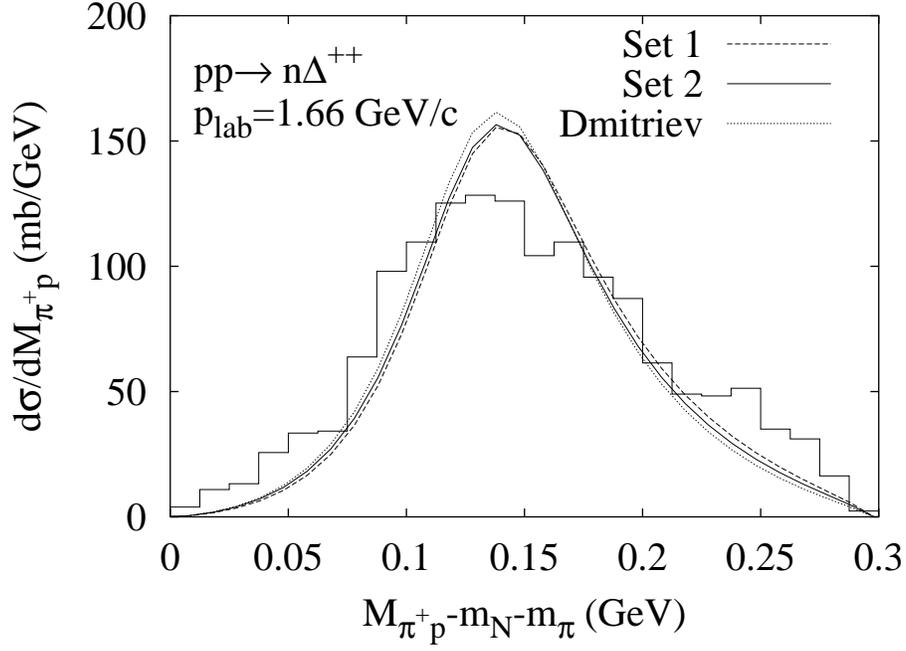}

\vspace{7cm}

\caption{\label{fig:sigma_vs_minv} Invariant mass distribution of 
outgoing $\pi^+$ and proton for $pp$ collisions at $p_{lab}=1.66$ GeV/c.
In the calculations only the $pp \to n\Delta^{++}$, 
$\Delta^{++} \to \pi^+ p$ reaction channel is taken into account.
The data from Ref.~\cite{Bugg64} are shown by the histogram.}
\end{figure}

Since our major goal is to describe the pion production 
from heavy-ion collisions at the beam energies $\sim$ 1 A GeV, we have 
also checked the differential cross sections $pp \to n p \pi^+$ at 
$p_{lab}=1.66$ GeV/c ($E_{lab} = 0.97$ GeV).
Fig.~\ref{fig:sigma_vs_costh} shows the c.m. polar angle distribution of 
neutrons. Our calculation is again in a very close agreement with 
the relativistic treatment of Ref. \cite{Dm86} and describes the data 
\cite{Bugg64} well. We neglected the contribution of the channel
$pp \to p\Delta^+~,~\Delta^+ \to \pi^+ n$ in comparison to the data: 
taking into account this channel would roughly multiply our results by
a factor of 10/9, which would yield an even better agreement with the data. 
The invariant mass distribution of $(\pi^+,p)$ pairs is shown in 
Fig.~\ref{fig:sigma_vs_minv}.
We see that both models agree quite well, but the experimental 
distribution from \cite{Bugg64} is somewhat broader. 
Notice, that the channel $pp \to p\Delta^+~,~\Delta^+ \to \pi^+ n$
missed in our calculations would create some broadening of the spectrum,
since now the pair ($\pi^+,p$) is not emitted from the same resonance. 
However, the study of this effect is beyond the scope of our work.

\subsection{ In-medium corrections }

\subsubsection{ Effective masses and in-medium $\Delta$ width}

\begin{figure}

\vspace{1cm}

\includegraphics{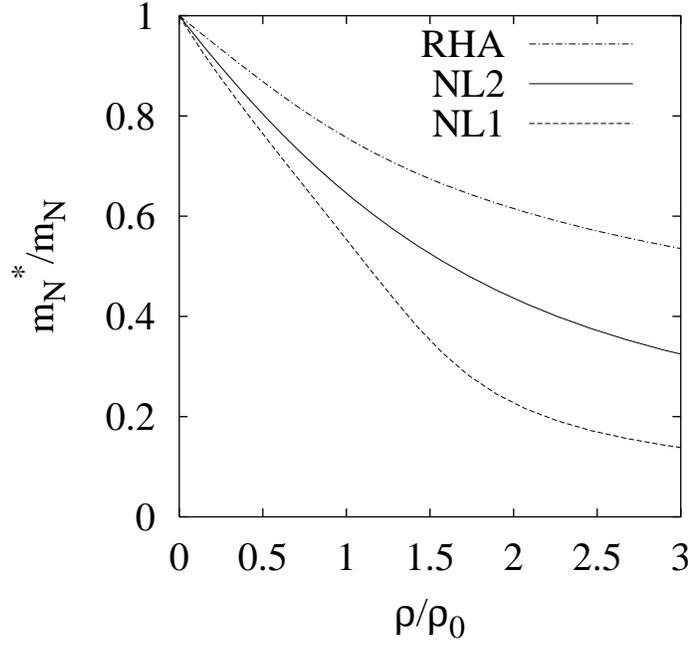}

\vspace{4cm}

\caption{\label{fig:waltst} Ratio of the effective nucleon mass
to the vacuum nucleon mass vs the ratio of the baryon density to
the nuclear saturation density. Dashed, solid and dash-dotted lines
show the calculations using NL1, NL2 and RHA respectively.}
\end{figure}

We assume that nucleons and $\Delta$'s are coupled to the scalar 
mean field $\sigma$ by the same universal coupling constant 
$g_\sigma$ \cite{Wehr89}. This produces the Dirac effective masses
\begin{equation}
    m_N^* = m_N + g_\sigma \sigma~,~~~
    m_\Delta^* = m_\Delta + g_\sigma \sigma~,           \label{mstar}
\end{equation}
which are smaller than the bare ones by 300$\div$400 MeV at the normal
nuclear density, as one can see from Fig.~\ref{fig:waltst}.
We have adopted the two parameter sets of the relativistic mean field model 
(RMF) \cite{Lee86}: NL1 and NL2, and the Relativistic Hartree Approximation 
(RHA) \cite{SW86}. RHA takes into account vacuum fluctuations which are 
missed in RMF. We observe from Fig.~\ref{fig:waltst}, that the RMF set NL1 
produces the steepest decrease of the nucleon effective mass with the baryon 
density, while RHA gives the slowest one.

Assuming, further, that the coupling constants to the vector mean field
$\omega$ are also the same for nucleons and $\Delta$'s \cite{Wehr89},
we get the kinetic four-momenta
\begin{equation}
    p_N^* = p_N - g_\omega \omega~,~~~
    p_\Delta^* = p_\Delta - g_\omega \omega~.           \label{kinmom}
\end{equation}
The effective masses and the kinetic four-momenta substitute the vacuum masses
and four-momenta in the calculations. Thus, in Eq. (\ref{sigvac}) one
should replace~:
\begin{eqnarray}
        &p_i& \rightarrow p_i^*~,~~~i=1,2,3,4~,           \\
        &m_N& \rightarrow m_N^*~,                        \label{mnstar}\\
        &M_\Delta& \rightarrow M_\Delta^*~,              \label{mdstar}\\
&I& \rightarrow I^* = \sqrt{(p_1^*p_2^*)^2 - m_N^{*4}}~,  \label{istar}\\
&{\cal A}_\Delta(M_\Delta^2)& \rightarrow 
 {\cal A}_\Delta^*(M_\Delta^{*2}) = {1 \over \pi}
        { M_\Delta^* \Gamma_\Delta^*(M_\Delta^*) \over 
          (M_\Delta^{*2}-m_\Delta^{*2})^2 
         + M_\Delta^{*2} \Gamma_\Delta^{*2}(M_\Delta^*) }~.      
                                                         \label{spfunmed}
\end{eqnarray}

The total in-medium $\Delta$ width is approximated as follows~:
\begin{equation}
        \Gamma_\Delta^*(M_\Delta^*) = \Gamma_{sp} 
        {\rho \over \rho_0}
          + \Gamma_\Delta^0\left({q(M_\Delta^*,m_N^*,m_\pi) 
                            \over q(m_\Delta,m_N,m_\pi)}\right)^3
         {m_\Delta^* \over M_\Delta^*} 
         {\beta_0^2+q^2(m_\Delta^*,m_N^*,m_\pi) 
    \over \beta_0^2+q^2(M_\Delta^*,m_N^*,m_\pi)}~,       \label{gammed}
\end{equation}
where $\rho_0=0.16$ fm$^{-3}$ is the nuclear saturation density
and $\Gamma_{sp}$ is a constant.
The first term in Eq.(\ref{gammed}) is the $\Delta$-spreading width 
\cite{Hir79,OS87}, which describes the in-medium broadening of the 
$\Delta$ resonance due to the absorption and rescattering processes 
$\Delta N \to NN$, $\Delta N \to \Delta N$ and $\Delta N N \to N N N$.
In Ref. \cite{OS87} the many-body calculations of the $\Delta$ 
self-energy in nuclear matter have been performed which are in a
good agreement with the empirical $\Delta$-spreading potential of
Ref. \cite{Hir79}
\begin{equation}
        W_{sp} = {(23\pm5) - i(43\pm5)}{\rho \over \rho_0}~(\mbox{MeV})~.
                                                         \label{Wsp}
\end{equation}
Thus, the value of the constant $\Gamma_{sp}$ consistent with
Refs. \cite{Hir79,OS87} is
$\Gamma_{sp} = -2\mbox{Im}W_{sp}(\rho=\rho_0) = 80$ MeV.
In Ref. \cite{RW94} another value $\Gamma_{sp} = 20$ MeV
has been used in the calculation of the pion self-energy within the
$\Delta$-hole model. In Ref. \cite{Kim97} the $\Delta$-spreading width
has been calculated in the relativistic meson-nucleon model including 
effective masses given by the Walecka model. Only the $\Delta N \to NN$ 
contribution has been taken into account. Vertex corrections were
neglected in Ref. \cite{Kim97}, which resulted in rather big values
of the $\Delta$-spreading width in the calculation without effective mass
($\Gamma_{sp} = 160$ MeV). Including effective mass corrections,
the $\Delta$-spreading width of Ref. \cite{Kim97}, first, increases with
density at $\rho < \rho_0$ in agreement with $\Gamma_{sp} = 80$ MeV, but
than drops abruptly at higher densities due to the mean field effects.
We have checked that the variation of the parameter $\Gamma_{sp}$ within the 
range $0\div80$ MeV leads to the changes in the in-medium cross section
$NN \to N\Delta$ which are comparable with the ambiguity due to the choice
of the effective mass parameterization (NL1 or NL2). On the other hand, 
we need to keep $\Gamma_{sp} \neq 0$ in order to have numerically stable 
results in the calculations of the $\Delta$-hole Lindhard function 
(\ref{LindD1}). In the numerical results presented below we use 
$\Gamma_{sp} = 80$ MeV.

The second term in Eq.(\ref{gammed}) is the $\Delta \to N \pi$ width taking 
into account the effective masses of the nucleon and $\Delta$. 
In this term we have omitted the Pauli blocking of the decay nucleon. 
This is justified because in the real heavy-ion collision at the beam energy 
of 1 A GeV the nuclear matter at finite temperature of about 70 MeV is 
created, which reduces the Pauli blocking effect strongly with respect to the 
zero temperature nuclear matter.

In the c.m. frame of colliding nucleons (i.e. in the frame where
${\bf p}_1^* + {\bf p}_2^* = 0$) the differential cross section in 
nuclear medium reads as follows (c.f. (\ref{sigvaccm}))~:
\begin{equation}
        {d\sigma^{med} \over dM_\Delta^{*2} d\Omega}
   = { (2m_N^*)^3 2M_\Delta^* \over 64 \pi^2 s^*} \overline{|T^{med}|^2}
     { q(\sqrt{s^*},M_\Delta^*,m_N^*) 
 \over q(\sqrt{s^*},m_N^*,m_N^*)} {\cal A}_\Delta^*(M_\Delta^{*2})~,
                                                         \label{sigmedcm}
\end{equation}
where $s^* = (p_1^* + p_2^*)^2$.
Below we will assume that the c.m. frame of colliding
particles coincides with the nuclear matter rest frame, where
spatial components of the vector field disappear, thus 
${\bf p}^* = {\bf p}$.

\subsubsection{ Pion self-energy }

The in-medium matrix element $T^{med}$ in Eq. (\ref{sigmedcm}) 
is given by the graphs shown in Fig.~\ref{fig:nnnd_opem}, where now the 
in-medium pion propagator
\begin{equation}
        D(k) = {1 \over k^2 - m_\pi^2 - \Pi(k) }         \label{piprop}
\end{equation}
must be put on the place of the vacuum propagator $D^{(0)}(k)$ and
the in-medium vertex corrections due to the contact interactions
must be also taken into account. The pion self-energy
$\Pi_{ab}(k)$ is, in general, a matrix with two isospin indices 
$a,b = 1,2,3$. However, in this Section we consider isospin-symmetric 
nuclear matter, where $\Pi_{ab}(k)=\delta_{ab}\Pi(k)$. 
The delta function of the isospin indices will be always dropped below for 
brevity. The assumption of isospin-symmetric nuclear matter results,
in particular, in the validity of the vacuum relations between various
isospin channels (c.f Eqs.(\ref{isorel1}),(\ref{isorel2}) below) for
the in-medium cross sections. To avoid misunderstanding, we stress, however,
that in the BUU calculations of Sect. III we take into account
the isospin assymmetry of colliding nuclei explicitly.

\begin{figure}

\vspace{1cm}

\includegraphics{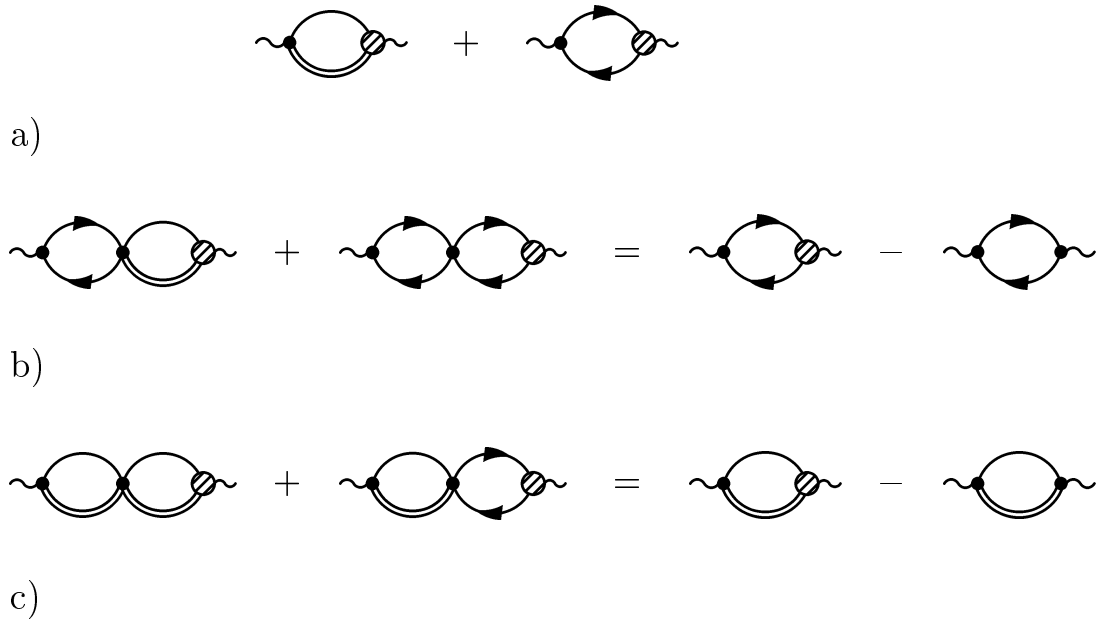}

\vspace{5cm}

\caption{\label{fig:nnnd_pipol} (a) -- Pion self-energy 
$-i\Pi(k)$. (b),(c) -- The recurrence
relations for the $\Delta N^{-1}$ and $N N^{-1}$ iterated contributions 
(see Eqs.(\ref{sys1})). Arrows are not shown on the $\Delta N^{-1}$ loops,
since each of these loops is the sum $s$- and $u$-channel graphs
(with anticlockwise and clockwise directed arrows, respectively).}
\end{figure}

The function $-i\Pi(k) = -i\Pi_\Delta(k) -i\Pi_N(k)$
is shown in Fig.~\ref{fig:nnnd_pipol}a. It includes the $\Delta$-hole and 
nucleon-hole iterated contributions: we denoted them as $-i\Pi_\Delta(k)$ 
and $-i\Pi_N(k)$ 
respectively. These contributions satisfy the recurrence relations shown in 
Fig.~\ref{fig:nnnd_pipol}b,c which read as~:
\begin{equation}
         \hat A(k) \left( \begin{array}{c}
                               \Pi_N(k) \\
                               \Pi_\Delta(k)
                       \end{array}
                \right)
       =   -{\bf k}^2 \left( \begin{array}{c}
                                     \chi_N(k) \\
                                     \chi_\Delta(k)
                             \end{array}
                      \right)~,                          \label{sys1}
\end{equation}
where the 2$\times$2 matrix $\hat A(k)$ is
\begin{equation}
        \hat A(k) \equiv \left( \begin{array}{cc}
1+g_{NN}^\prime\chi_N(k) & g_{N\Delta}^\prime\chi_N(k) \\
g_{N\Delta}^\prime\chi_\Delta(k) & 1+g_{\Delta\Delta}^\prime\chi_\Delta(k)
                        \end{array}
                 \right)~.                                \label{Amatr}
\end{equation}
The susceptibilities $\chi_N(k)$ and $\chi_\Delta(k)$ are~:
\begin{eqnarray}
  \chi_N(k) & = & {4 f^2(k^2) \over m_\pi^2 } \phi(k)~, \label{chin} \\
  \chi_\Delta(k) & = & {16 f_\Delta^2(k^2) \over 9 m_\pi^2 }
    ( \phi_\Delta(k) + \phi_\Delta(-k) )                \label{chid}
\end{eqnarray}
with $\phi(k)$ and $\phi_\Delta(\pm k)$ being, respectively, the nucleon-hole 
and $\Delta$-hole Lindhard functions~:
\begin{eqnarray}
\phi(k) & = & i \int {d^4 p \over (2\pi)^4} G(p) G(p+k)~,  \label{LindN} \\
\phi_\Delta(\pm k) & = & i \int {d^4 p \over (2\pi)^4}
                           G(p) G_\Delta(p \pm k)~.         \label{LindD}
\end{eqnarray}
Here $G(p)$ and $G_\Delta(p_\Delta)$ are the nucleon and $\Delta$ 
propagators~:
\begin{eqnarray}
G(p) & = & {1 \over p^0 - \varepsilon^*({\bf p}) + i0} +
2 \pi i\, n({\bf p}) \delta( p^0 - \varepsilon^*({\bf p}) )~, 
                                                           \label{nucprop} \\
G_\Delta(p_\Delta) & = & 
{ 1 \over p_\Delta^0 - \varepsilon_\Delta^*({\bf p}_\Delta)
  + i\Gamma_\Delta^*(\sqrt{p_\Delta^2})/2 }~,              \label{delprop}
\end{eqnarray}
where 
\begin{equation}
\varepsilon^*({\bf p}) = \sqrt{{\bf p}^2 + (m_N^*)^2}~,~~~
\varepsilon_\Delta^*({\bf p}_\Delta) =
              \sqrt{{\bf p}_\Delta^2 + (m_\Delta^*)^2}      \label{enonsh}
\end{equation}
are the on-shell nucleon and $\Delta$ energies, $n({\bf p})$ is 
the nucleon occupation number.
The nucleon propagator (\ref{nucprop}) consists of the vacuum part and
the in-medium part ($\propto n({\bf p})$). The $\Delta$ propagator 
(\ref{delprop}) includes the vacuum part only, since we have neglected 
the presence of $\Delta$ excitations in nuclear matter. Both propagators
take into account the effective mass corrections. 

After the contour integration over $p^0$ (c.f. \cite{FW71}) the 
nucleon-hole Lindhard function (\ref{LindN}) takes the following form~:
\begin{equation}
   \phi(k) = - \int {d^3 p \over (2\pi)^3} \left(
   { n({\bf p}+{\bf k})(1-n({\bf p})) \over
     \varepsilon^*({\bf p}+{\bf k}) - k^0 - \varepsilon^*({\bf p}) + i0 }
+  { n({\bf p})(1-n({\bf p}+{\bf k})) \over
     \varepsilon^*({\bf p}) + k^0 - \varepsilon^*({\bf p}+{\bf k}) + i0 }
                                           \right)~.    
                                                      \label{LindN1}
\end{equation}
We have used the analytic formulas given in Ref. \cite{FW71} for the 
calculation of the integral (\ref{LindN1}) in the case of Fermi distribution 
at zero temperature $n({\bf p})=\Theta(p_F-|{\bf p}|)$.

The $\Delta$-hole Lindhard function (\ref{LindD}), after integration over
$p^0$, can be expressed as
\begin{equation}
  \phi_\Delta(\pm k) = - \int {d^3 p \over (2\pi)^3}
   { n({\bf p}) \over \varepsilon^*({\bf p}) \pm k^0
     - \varepsilon_\Delta^*({\bf p} \pm {\bf k})
     + i \Gamma_\Delta^*(\sqrt{(p^* \pm k)^2})/2 }~,      \label{LindD1}
\end{equation}
where $p^* \equiv (\varepsilon^*({\bf p}),{\bf p})$. In the calculation
of the integral (\ref{LindD1}) we have used the expression from the 
Appendix of Ref. \cite{OP81} (see also the Appendix in Ref. \cite{Oset90})
assuming that the invariant mass of the nucleon and pion in the
argument of $\Gamma^*$ can be approximated by
\begin{equation}
   (p^* \pm k)^2 \simeq (m_N^*)^2 + k^2 
   \pm 2 k^0 \sqrt{(m_N^*)^2 + {3 \over 5} p_F^2}~.        \label{appr}
\end{equation}

Solving the system (\ref{sys1}) we get for the pion self-energy
the following expression~:
\begin{eqnarray}
        &\Pi(k)& = \Pi_\Delta(k) + \Pi_N(k)           
= -{\bf k}^2 {\det}^{-1}(\hat A(k))                             \nonumber \\  
&\times& [ (1+g_{\Delta\Delta}^\prime\chi_\Delta(k))\chi_N(k)
+ (1+g_{NN}^\prime\chi_N(k))\chi_\Delta(k)
- 2 g_{N\Delta}^\prime \chi_N(k) \chi_\Delta(k) ]~,        \label{pipol}
\end{eqnarray}
where
\begin{equation}
     \det(\hat A(k)) = (1+g_{NN}^\prime\chi_N(k)) 
(1+g_{\Delta\Delta}^\prime\chi_\Delta(k)) 
-  (g_{N\Delta}^\prime)^2 \chi_N(k) \chi_\Delta(k)~.       \label{DetA}
\end{equation}
Expression (\ref{pipol}) coincides with the result from 
Ref. \cite{Xia88}.

\subsubsection{ Vertex corrections }

\begin{figure}

\vspace{1cm}

\includegraphics{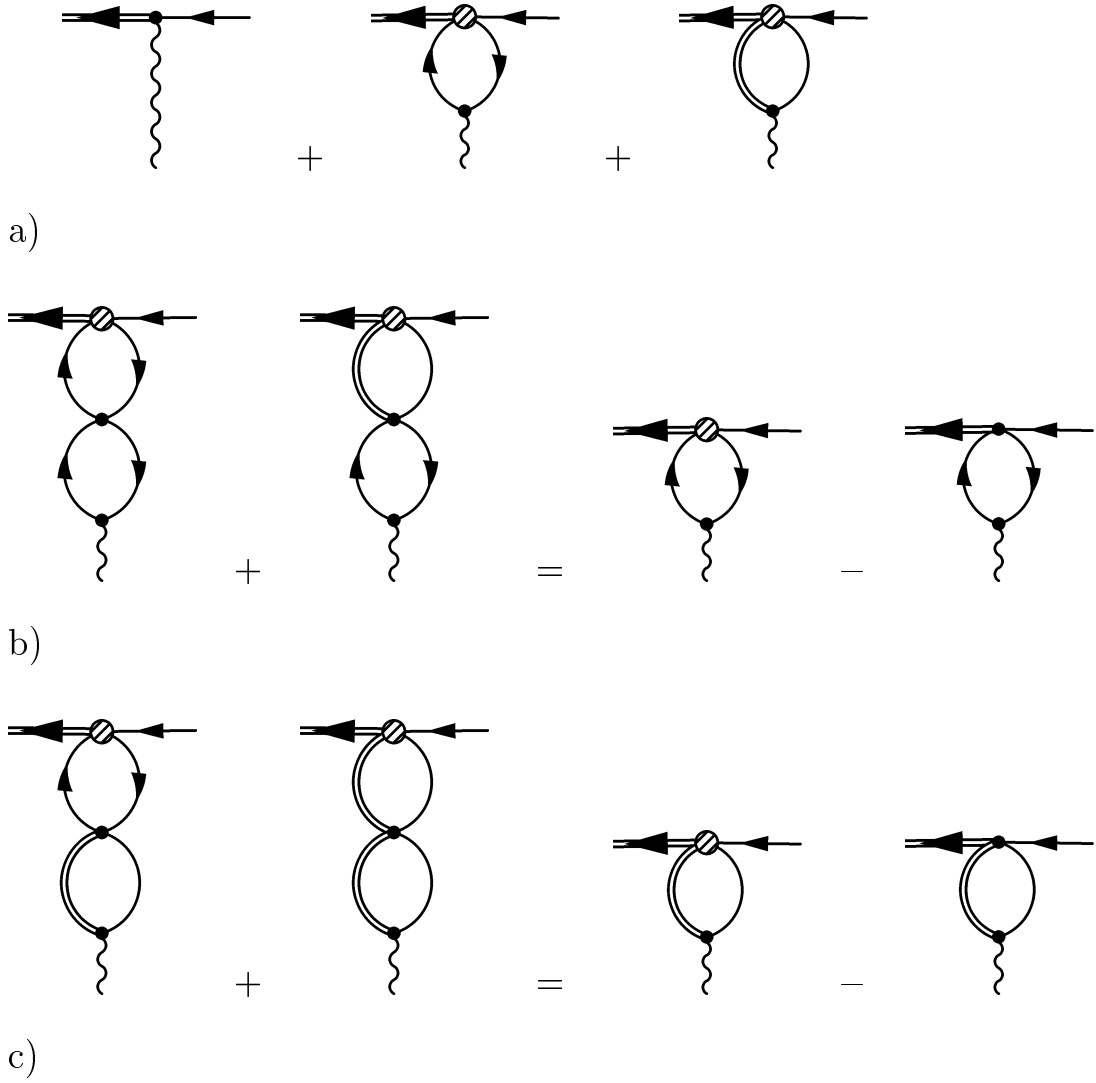}

\vspace{4cm}

\caption{\label{fig:nnnd_pind} (a) -- Renormalized $\pi N \Delta$ vertex
operator $\tilde\Gamma_j^{abs}(k)$. (b),(c) --  Recurrence relations
for the $\Delta N^{-1}$ and $N N^{-1}$ iterated contributions 
to $\tilde\Gamma_j^{abs}(k)$ (see Eqs.(\ref{sys2})).} 
\end{figure}

The renormalized $\pi N \to \Delta$ absorption vertex operator 
$\tilde\Gamma_j^{abs}(k)$ is shown in Fig.~\ref{fig:nnnd_pind}a. 
The first graph in Fig.~\ref{fig:nnnd_pind}a is the $\pi N \to \Delta$
absorption vertex operator in vacuum $\Gamma_j^{abs}(k)$
(see Eq.(\ref{gama0})). The second and third graphs 
in Fig.~\ref{fig:nnnd_pind}a correspond to the iterated nucleon-hole and 
$\Delta$-hole corrections which we will denote as 
$\tilde\Gamma_{j,N}^{abs}(k)$ and $\tilde\Gamma_{j,\Delta}^{abs}(k)$ 
respectively.
The correction terms satisfy the recurrence relations shown in 
Fig.~\ref{fig:nnnd_pind}b,c which can be written as~:
\begin{equation}
         \hat A(k) \left( \begin{array}{c}
                               \tilde\Gamma_{j,N}^{abs}(k) \\
                               \tilde\Gamma_{j,\Delta}^{abs}(k)
                           \end{array}
                   \right)
       =   -\Gamma_j^{abs}(k) \left( \begin{array}{c}
                                g_{N\Delta}^\prime \chi_N(k) \\
                                g_{\Delta\Delta}^\prime\chi_\Delta(k)
                                     \end{array}
                              \right)~.                       \label{sys2}
\end{equation}
After solution of the system (\ref{sys2}) we obtain the expression
for the operator $\tilde\Gamma_j^{abs}(k)$~:
\begin{eqnarray}
   &\tilde\Gamma_j^{abs}(k)& = \Gamma_j^{abs}(k) 
 + \tilde\Gamma_{j,N}^{abs}(k) + \tilde\Gamma_{j,\Delta}^{abs}(k) \nonumber \\
&=& {\det}^{-1}(\hat A(k)) 
    [1 + (g_{NN}^\prime-g_{N\Delta}^\prime)\chi_N(k)] 
    \Gamma_j^{abs}(k)~.                                       \label{gama}
\end{eqnarray}

\begin{figure}

\vspace{1cm}

\includegraphics{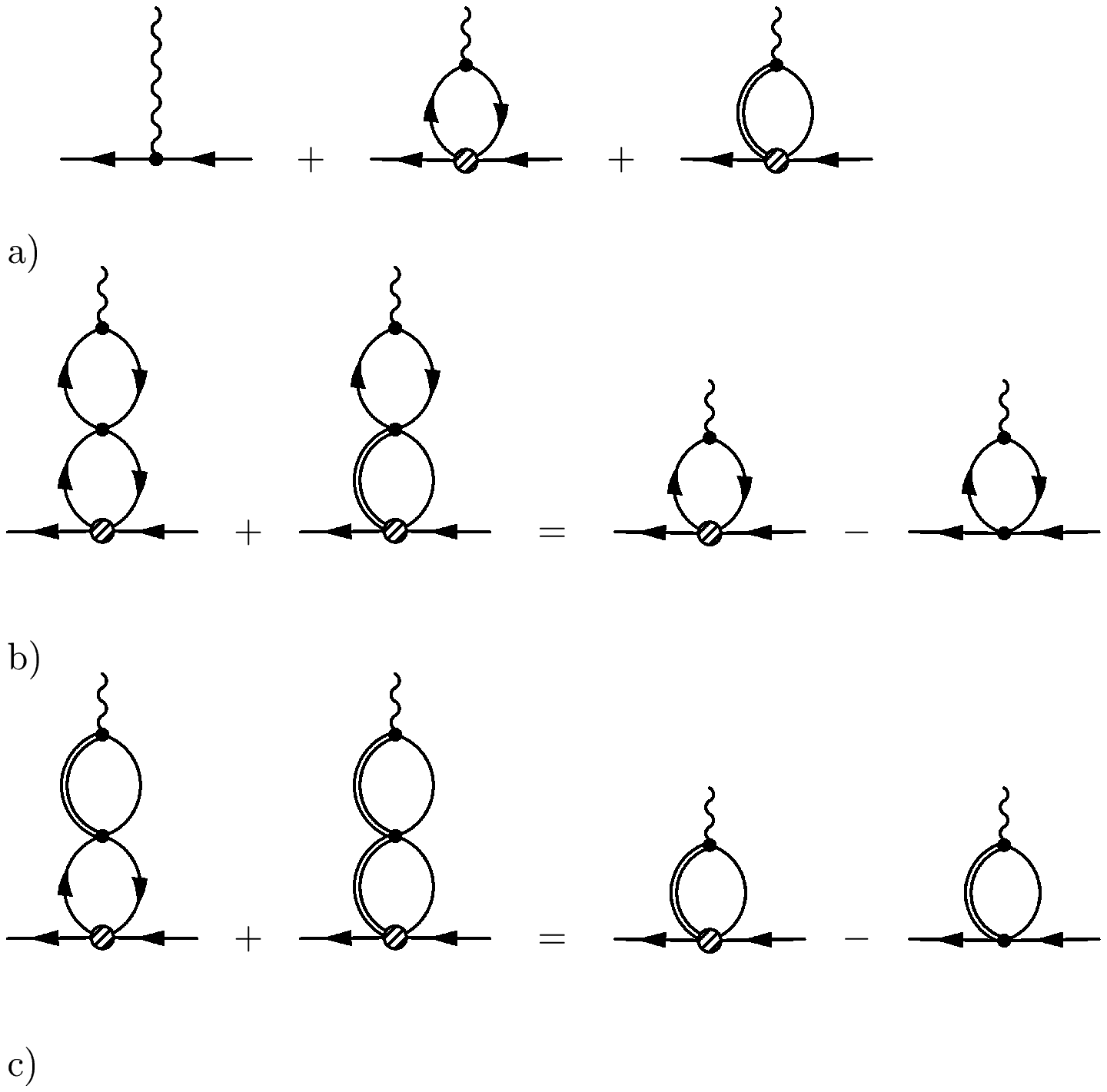}

\vspace{4cm}

\caption{\label{fig:nnnd_pinn} (a) -- Renormalized $\pi N N$ vertex
operator $\tilde\Gamma_j^{dec}(k)$. (b),(c) --  Recurrence relations
for the $\Delta N^{-1}$ and $N N^{-1}$ iterated contributions 
to $\tilde\Gamma_j^{dec}(k)$ (see Eqs.(\ref{sys3})).}
\end{figure}

In an analogous way, the renormalized $N \to \pi N$ decay vertex operator
$\tilde\Gamma_j^{dec}(k)$ can be represented as a sum of the vacuum
operator $\Gamma_j^{dec}(k)$, the iterated nucleon-hole 
$\tilde\Gamma_{j,N}^{dec}(k)$ and $\Delta$-hole 
$\tilde\Gamma_{j,\Delta}^{dec}(k)$ contributions (see 
Fig.~\ref{fig:nnnd_pinn}a).
The recurrence relations for $\tilde\Gamma_{j,N}^{dec}(k)$ and
$\tilde\Gamma_{j,\Delta}^{dec}(k)$ are shown in Fig.~\ref{fig:nnnd_pinn}b,c.
Algebraically, they look as~:
\begin{equation}
         \hat A(k) \left( \begin{array}{c}
                               \tilde\Gamma_{j,N}^{dec}(k) \\
                               \tilde\Gamma_{j,\Delta}^{dec}(k)
                           \end{array}
                   \right)
       =   -\Gamma_j^{dec}(k) \left( \begin{array}{c}
                                g_{NN}^\prime \chi_N(k) \\
                                g_{N\Delta}^\prime\chi_\Delta(k)
                                     \end{array}
                              \right)~.                       \label{sys3}
\end{equation}
Resolving (\ref{sys3}) one gets the renormalized vertex operator
$\tilde\Gamma_j^{dec}(k)$~:
\begin{eqnarray}
   &\tilde\Gamma_j^{dec}(k)& = \Gamma_j^{dec}(k) 
 + \tilde\Gamma_{j,N}^{dec}(k) + \tilde\Gamma_{j,\Delta}^{dec}(k) \nonumber \\
&=& {\det}^{-1}(\hat A(k)) 
    [1 + (g_{\Delta\Delta}^\prime-g_{N\Delta}^\prime)\chi_\Delta(k)] 
    \Gamma_j^{dec}(k)~.                                       \label{gamd}
\end{eqnarray}

\begin{figure}

\vspace{1cm}

\includegraphics{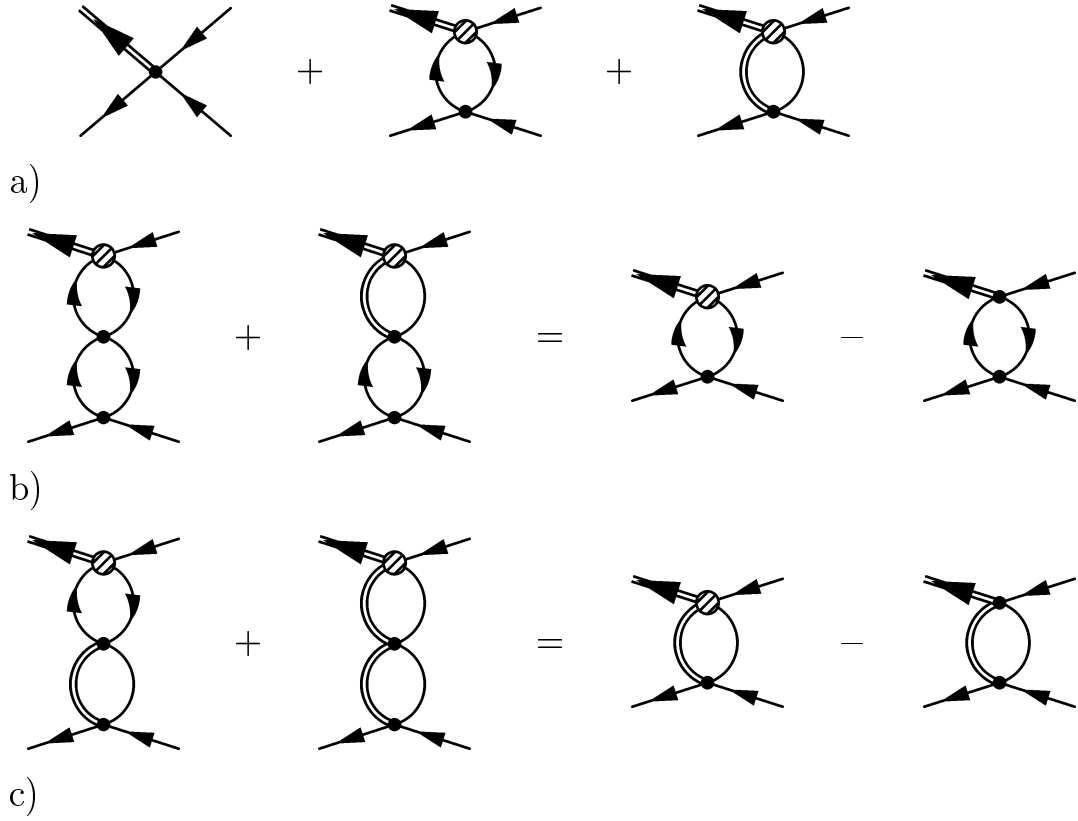}

\vspace{4cm}

\caption{\label{fig:nnnd_vcont} (a) -- Renormalized contact interaction
$-i \tilde{V}(k)$. (b),(c) -- Recurrence relations for the $\Delta N^{-1}$ 
and $N N^{-1}$ iterated contributions 
to  $-i \tilde{V}(k)$ (see Eqs.(\ref{sys4})).}
\end{figure}

Finally, Fig.~\ref{fig:nnnd_vcont}a shows the renormalized contact interaction 
$-i \tilde{V}(k)$ (we drop here the indices of external particles for
brevity) given by the sum of the bare contact interaction $-i V(k)$,
nucleon-hole $-i \tilde{V}_N(k)$ and $\Delta$-hole  
$-i \tilde{V}_\Delta(k)$ iterated contributions. The recurrence relations
for $-i \tilde{V}_N(k)$ and $-i \tilde{V}_\Delta(k)$, shown graphically
in Fig.~\ref{fig:nnnd_vcont}b,c, are~:
\begin{equation}
\left( \begin{array}{cc}
1+g_{NN}^\prime\chi_N(k) & g_{NN}^\prime \chi_N(k)\\
{(g_{N\Delta}^\prime)^2 \over g_{NN}^\prime}  \chi_\Delta(k) & 
1+g_{\Delta\Delta}^\prime\chi_\Delta(k)
       \end{array}\right)         
\left( \begin{array}{c}
          \tilde{V}_N(k) \\
          \tilde{V}_\Delta(k)
                           \end{array}
                   \right)
       =   -V(k) \left( \begin{array}{c}
                           g_{NN}^\prime \chi_N(k) \\
                           g_{\Delta\Delta}^\prime\chi_\Delta(k)
                        \end{array}
                 \right)~.                                   \label{sys4}
\end{equation}
Resolving the system (\ref{sys4}) we get a simple formula for
$\tilde{V}(k)$~:
\begin{equation}
        \tilde{V}(k) = V(k) + \tilde{V}_N(k) + \tilde{V}_\Delta(k)
  = {\det}^{-1}(\hat A(k)) V(k)~.                            \label{Vcont}
\end{equation}

The in-medium matrix element $T^{med}$ is given by the sum of the 
direct Eq.(\ref{Ta}) and exchange Eq.(\ref{Tb}) contributions where
$\Gamma$ and $V$ are replaced by the renormalized quantities.
The explicit formulas for calculation of $\overline{|T^{med}|^2}$ are
given in Appendix A.

\subsubsection{ Cross sections }

Fig.~\ref{fig:sigma_vs_ekcm_rho} shows the total in-medium 
$pp \to n \Delta^{++}$ cross section vs the c.m. kinetic energy above 
the pion production threshold
\begin{equation}
         Q^* = \sqrt{s^*} - 2m_N^* - m_\pi            \label{Qstar}
\end{equation}
for different densities.
We compare the cross sections at fixed $Q^*$
rather than at fixed c.m. momentum, because $Q^*$ is the direct
measure of the collision inelasticity: 
e.g., $\sigma_{pp \to n \Delta^{++}}^{med}(Q^* = 0) = 0$ for all baryon 
densities, while the threshold value of the c.m. momentum is density 
dependent. The solid lines in the panels a,b and c of 
Fig.~\ref{fig:sigma_vs_ekcm_rho} show the results of the full calculation
(i.e. including effective masses, the pion collectivity and vertex 
renormalization) for the combinations NL1-Set1, NL1-Set2 and NL2-Set2,
respectively. We see that the cross section quickly drops with baryon 
density. The strongest effect is obtained for the combination NL1-Set2: 
in this case $\sigma_{pp \to n \Delta^{++}}^{med}$ drops by two orders 
of magnitude when the density increases from 0 up to $3 \rho_0$. 
The combination NL1-Set1 produces a somewhat weaker decrease of 
$\sigma_{pp \to n \Delta^{++}}^{med}$ with density than the 
combination NL1-Set2. However, an effect of the choice of the 
OPEM parameters on the result is much smaller than the effect of 
the choice of the effective nucleon and $\Delta$ masses. The set NL2
gives less in-medium reduction than the set NL1 as one expects 
from Fig.~\ref{fig:waltst}. 

The main reason for the in-medium reduction of the cross section is the 
factor of $(2m_N^*)^3 2M_\Delta^*/s^*$ in front of the matrix 
element squared in Eq. (\ref{sigmedcm}) for the differential cross section. 
To demonstrate this, we have repeated the calculation using the NL2-Set2 
combination replacing this factor by its vacuum value, i.e. 
$(2m_N^*)^3 2M_\Delta^*/s^* \to  (2m_N)^3 2M_\Delta/s$ (see dashed lines in 
the panel c of Fig.~\ref{fig:sigma_vs_ekcm_rho}). One observes that the 
cross section calculated with the vacuum factor practically does not depend 
on the density. Further replacements of the c.m. momenta of the 
incoming and outgoing particles by their vacuum values and also of the 
effective by vacuum masses in the spectral function in Eq. (\ref{sigmedcm}) 
have no any visible effect.
 
In order to better understand the reason for the lack of the density 
dependence in the calculation with the vacuum  phase space factor
we have also calculated the cross section using the Set2 of the OPEM parameters
with vacuum masses both in the matrix element and phase space factors
(panels e and f of Fig.~\ref{fig:sigma_vs_ekcm_rho}). Panels e and f
of Fig.~\ref{fig:sigma_vs_ekcm_rho} show the results with and without
pion collectivity and vertex renormalization effects, respectively.
Namely, in the calculation presented in panel f we have put all the Lindhard
functions equal to zero. However, the $\Delta$-spreading width is still 
present in the results shown in panels e and f. 
(i) As one can see from panel e, the cross section drops with
density even without any inclusion of the effective masses. By comparing
the solid lines in panel e with the dashed lines in panel c we conclude that 
the inclusion of the effective masses in the matrix element increases the 
cross section. 
(ii) From panel f we observe that the $\Delta$-spreading width in the 
spectral function reduces the cross section slightly. As discussed above,
this effect is comparable with the ambiguity given by the choice of the 
effective mass parameterization. 
(iii) By comparing the curves in panels e and f one sees that the pion 
collectivity and vertex renormalization reduce the cross section rather
noticeably.
 Therefore, the density independence observed for the calculation 
with the vacuum phase space factor only (dashed lines in panel c) is,
basically, due to a counterbalance of the effects (i) and (iii).

Panel d of Fig.~\ref{fig:sigma_vs_ekcm_rho} shows the NL2-Set2 calculation
with the Lindhard functions putted to zero in the matrix element (solid lines)
and also the calculation using the matrix element of Ref. \cite{Dm86} 
with the in-medium nucleon and $\Delta$ masses from the NL2 set
(dashed lines). As expected, both calculations are in a very close agreement
since the nucleon-hole and $\Delta$-hole Lindhard functions
are absent in both cases.  

\begin{figure}

\vspace{1cm}

\includegraphics{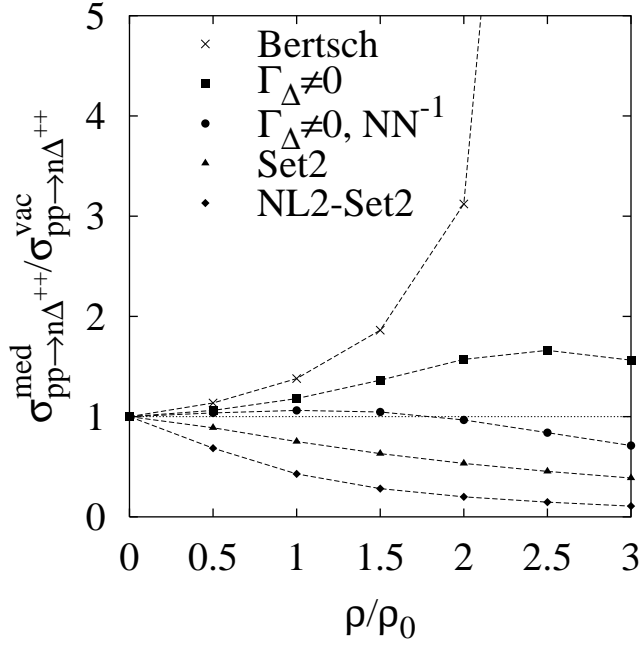}

\vspace{4cm}

\caption{\label{fig:sigma_vs_rho} Density dependence of the ratio of 
the in-medium and vacuum total cross sections $pp \to n\Delta^{++}$ for 
$Q^* = 0.225$ GeV (in vacuum this corresponds to the proton beam energy 
of 0.8 GeV). Line marked with crosses corresponds to the approximations
of Ref. \cite{Bertsch88} with the parameter set B \cite{Bertsch88}.
Boxes and circles show the calculations including the in-medium
$\Delta$ width without and with the $NN^{-1}$ Lindhard function 
contribution respectively. Triangles and rombuses show the results
for the Set2 of the OPEM parameters without and with inclusion 
of the effective mass, respectively.}
\end{figure}

A comment is in order here on the relation to the results of 
Ref. \cite{Bertsch88} where it was shown neglecting effective mass 
corrections that the in-medium cross section 
$\sigma_{pp \to n \Delta^{++}}^{med}$ grows with baryon density.
We have repeated the calculations of \cite{Bertsch88} with the parameter
set B ($\Lambda=0.565$ GeV, $g_{N\Delta}^\prime=1/3$) for the proton
beam energy $E_{lab}=0.8$ GeV. Following Ref. \cite{Bertsch88} we
have kept only the $\Delta$-hole Lindhard function in the quasistatic
approximation with zero width of the $\Delta$ resonance 
\begin{equation}
       \chi_\Delta(k) \simeq - {8f_\Delta^2(k^2) \over 9m_\pi^2}
                               \rho {\omega_\Delta \over 
                                     (k^0)^2 - \omega_\Delta^2}~,
                                                            \label{chiqsa}
\end{equation}
where $\omega_\Delta = m_\Delta - m_N + {\bf k}^2/(2m_\Delta)$.
In the $\Delta$ spectral function the $\Delta$ width has been also
put to zero \cite{Bertsch88}. Within these approximations we have
reproduced the result of Ref. \cite{Bertsch88} on the in-medium
enhancement of $\sigma_{pp \to n \Delta^{++}}^{med}$ (see crosses in
Fig.~\ref{fig:sigma_vs_rho}). The inclusion of the in-medium $\Delta$ 
width reduces the in-medium enhancement (see filled boxes in 
Fig.~\ref{fig:sigma_vs_rho}), which has been 
also demonstrated  earlier in Ref. \cite{WuKo89}. We have found that
taking into account the nucleon-hole Lindhard function makes 
$\sigma_{pp \to n \Delta^{++}}^{med}$ a decreasing function of the
density at $\rho > \rho_0$ in the calculations with vacuum masses 
(filled circles in Fig.~\ref{fig:sigma_vs_rho}). In the calculations with 
effective masses the nucleon-hole Lindhard function has a very small effect.
For the choice of the Landau-Migdal parameters and of the
cut-off factor from Set2 the cross section decreases faster with density 
(triangles in Fig.~\ref{fig:sigma_vs_rho}). Introducing the effective masses 
amplifies the in-medium reduction effect strongly (see rombuses in 
Fig.~\ref{fig:sigma_vs_rho}).

Summarizing all these findings, we can state that by just changing the 
factor in front of the matrix element in (\ref{sigvaccm}) to its
in-medium value takes care of most of the in-medium corrections on
the $\Delta$-production cross section. This finding offers a way that
can easily be implemented into any numerical realization.

\begin{figure}

\vspace{1cm}

\includegraphics{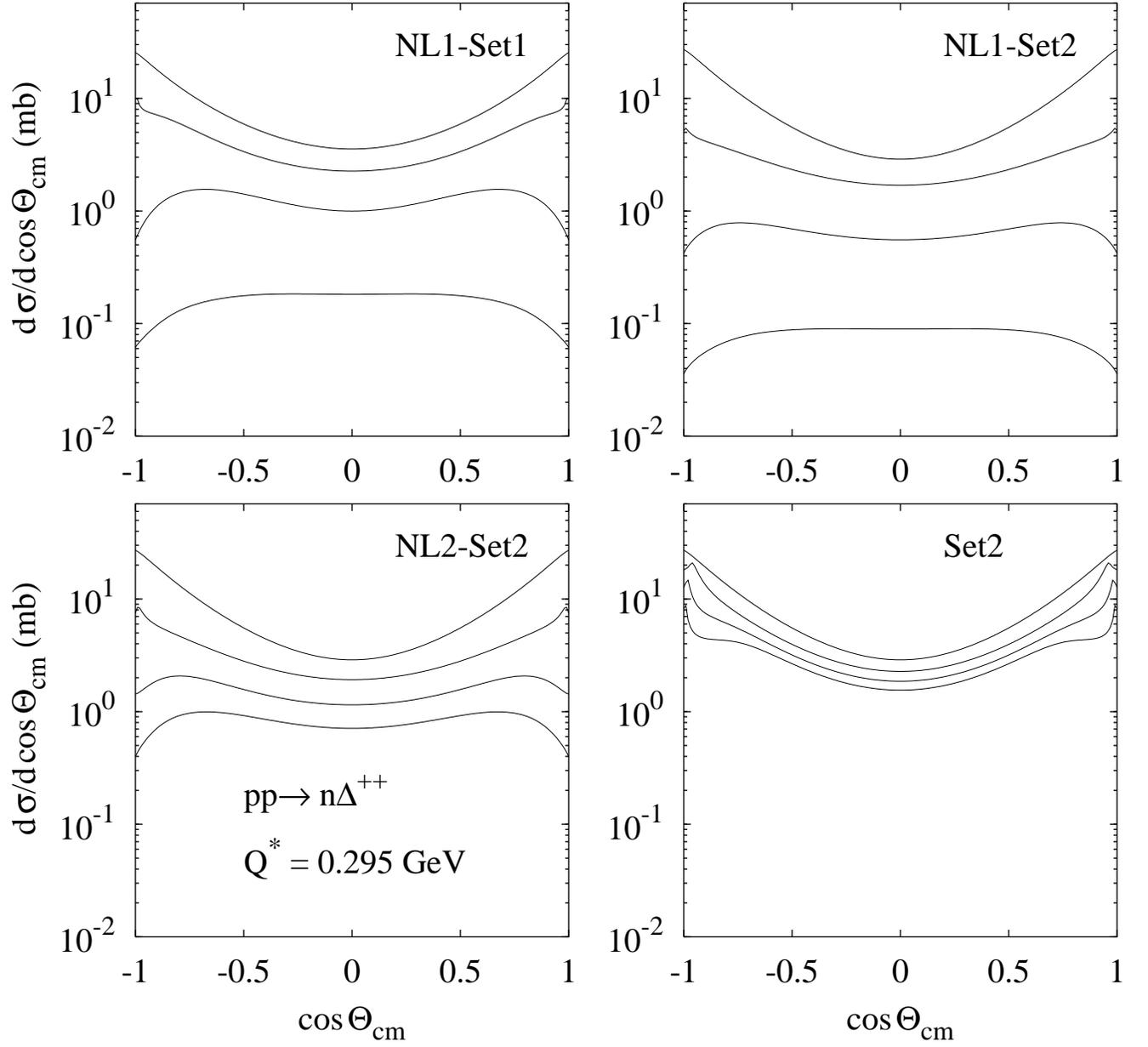}

\vspace{4cm}

\caption{\label{fig:dsigdcos} Neutron c.m. polar angle dependence of the 
in-medium differential $pp \to n\Delta^{++}$ cross section. The lines
from the uppermost to the lowermost correspond to the baryon density
$\rho = 0,~\rho_0,~2\rho_0$ and $3\rho_0$, respectively. Calculations
are performed for $Q^* = 0.295$ GeV.}
\end{figure}

In Fig.~\ref{fig:dsigdcos} we show the in-medium differential 
cross section of the outgoing neutron at various densities 
for $Q^* = 0.295$ GeV which corresponds to $p_{lab} = 1.66$ GeV/c at 
$\rho = 0$. For the calculation without effective mass (lower right panel)
the shape of the angular distribution does not depend, practically, on 
the density for $60^o < \Theta_{cm} < 120^o$: only changes for 
forward/backward angles are visible. However, in the calculations with
effective masses (all other pannels) the cross section becomes more 
isotropic with increasing density in qualitative agreement with the 
Dirac-Brueckner calculations of the elastic NN cross section in nuclear 
matter \cite{Li93,Fuchs01}. 

\section{ BUU calculations }

The in-medium modified $p p \to n \Delta^{++}$ cross section has been
implemented in the BUU program of the version described in 
Ref. \cite{EBM99}. More details can be found in Ref. \cite{EffePhD}.  
The cross sections for another isospin channels are related to the
$p p \to n \Delta^{++}$ cross section by the Clebsch-Gordan coefficients
which gives:
\begin{eqnarray}
        &\sigma_{pp \to p\Delta^+}^{med}& 
       = \sigma_{pn \to p\Delta^0}^{med} 
       = \sigma_{pn \to n\Delta^+}^{med}                      
       =\sigma_{nn \to n\Delta^0}^{med}
       = {1 \over 3} \sigma_{pp \to n\Delta^{++}}^{med}~,  \label{isorel1} \\
        &\sigma_{nn \to p\Delta^-}^{med}&
       = \sigma_{pp \to n\Delta^{++}}^{med}~.              \label{isorel2}
\end{eqnarray}

Besides the $\Delta$-excitation, there are another channels of the pion
production in a NN collision. At collision energies above the pion
production threshold $\sqrt{s} > 2m_N + m_\pi$ not only the $\Delta(1232)$,
but any other resonance $R$ can be excited~:
\begin{equation}
       N N \to N R~,                                         \label{NNNR}
\end{equation}
where $R$ stands for N$^*$(1440),  N$^*$(1535), N$^*$(1650) ...
(see Ref. \cite{EBM99} for the full list of resonances implemented
in the BUU code). At $\sqrt{s} > 2(m_N + m_\pi)$ the double $\Delta$
production channel opens~:
\begin{equation}
       N N \to \Delta \Delta~.                               \label{NNDD}
\end{equation}
In addition, at collision energies slightly above the pion production threshold
the $s$-wave direct pion production mechanism is important~:
\begin{equation}
       N N \to N N \pi~.                                     \label{NNNNpi}
\end{equation}

In analogy with the density modification of the $\Delta$ resonance
cross section, we have also modified the processes (\ref{NNNR}),(\ref{NNDD})
and (\ref{NNNNpi}) keeping, however, the constant density independent matrix
elements as determined in Ref. \cite{Teis97}.

For the in-medium cross section for the process (\ref{NNNR}) we apply
Eq. (\ref{sigmedcm}), where any higher resonance replaces the
$\Delta$ resonance and $\overline{|T^{med}|^2}$ is now a constant depending
on the isospin channel. The effective mass of a higher resonance $R$
is supposed to be
\begin{equation}
       M_R^* = M_R + g_\sigma \sigma~,                       \label{MRstar}
\end{equation}
i.e. we assume the same mass shift for all baryon resonances.
The spectral function of a higher resonance $R$ is given by Eq. 
(\ref{spfunmed}) with replacement ``$\Delta$'' $\to$ ``$R$'', where
the medium-modified total width is a sum of partial widths~:
\begin{eqnarray}
   \Gamma_R^*(M_R^*) &=& \sum_i\, \Gamma_{R,i}^*(M_R^*)~, \label{GamRst} \\
   \Gamma_{R,i}^*(M_R^*) &=& \Gamma_{R,i}^0
   { \rho_i^*(M_R^*) \over \rho_i^*(m_R^*) }~,             \label{GamRist}
\end{eqnarray}
where $m_R^* = m_R + g_\sigma \sigma$. $m_R$ is the pole mass of a
resonance $R$ and 
\begin{eqnarray}
   \rho_i^*(M_R^*) &=& \int\, d(M_B^*)^2 d(M_m)^2 {\cal A}_B^*((M_B^*)^2)
                                                       \nonumber \\
   &\times& {\cal A}_m((M_m)^2) {q(M_R^*,M_B^*,M_m) \over M_R^*}
   B_{l_i}^2(q(M_R^*,M_B^*,M_m)/\beta_0)~.                 \label{rhoi}
\end{eqnarray}
Here, the lower indices ``$B$'' and ``$m$'' mean the outgoing baryon
and meson for the $i$-th decay channel: ($B,m$) = ($N,\pi$), ($N,\eta$),
($N,\omega$), ($\Lambda,K$), ($\Delta(1232),\pi$), ($N,\rho$),
($N,\sigma$), ($N(1440),\pi$), ($\Delta(1232),\rho$).
${\cal A}_B^*((M_B^*)^2)$ and ${\cal A}_m((M_m)^2)$ are the spectral 
functions of the outgoing particles, $B_{l_i}^2$ is a Blatt-Weisskopf 
barrier penetration factor corresponding to the relative angular momentum 
$l_i$ of the outgoing baryon and meson. 
Eqs. (\ref{GamRst}),(\ref{GamRist}),(\ref{rhoi})
are the in-medium modification of the vacuum widths from Ref. \cite{EBM99}. 
We have to remark only, that also for the outgoing $\Lambda$ we take in 
Eq. (\ref{rhoi}) an in-medium spectral function 
${\cal A}_\Lambda^*((M_\Lambda^*)^2) 
\equiv \delta((M_\Lambda^*)^2 - (m_\Lambda + g_\sigma \sigma)^2)$.
This is nessary in order to keep the decay channel $R \to \Lambda K$
open in nuclear matter, since the resonance $R$ has now reduced mass.  
However, we did not modify any cross sections involving strange particles
which is out of scope of our work. 

For the in-medium cross section of the process $NN \to \Delta\Delta$
we apply the following formula~:
\begin{eqnarray}
        {d \sigma_{N_1N_2 \to \Delta_3\Delta_4}^{med} 
   \over d (M_3^*)^2 d (M_4^*)^2}&=&{ (2m_N^*)^2 (2m_\Delta^*)^2
                                \over 16 \pi s^*}
\overline{|T_{N_1N_2 \to \Delta_3\Delta_4}|^2} S_{34}  \nonumber \\
&\times& {q(\sqrt{s^*},M_3^*,M_4^*) \over q(\sqrt{s^*},m_N^*,m_N^*)}
{\cal A}_\Delta^*((M_3^*)^2) {\cal A}_\Delta^*((M_4^*)^2)~,
                                              \label{sigmedDDcm}
\end{eqnarray}
where the matrix element $\overline{|T_{N_1N_2 \to \Delta_3\Delta_4}|^2}$
depends on the isospins of incoming and outgoing particles only.
$S_{ij} = 1~(1/2)$ if the i-th and j-th particles are in the different
(same) isospin states. In the vacuum the cross section (\ref{sigmedDDcm})
is the same as in Ref. \cite{EffePhD}.

Finally, we have also introduced effective mass modifications 
for the process of the direct $s$-wave pion production $NN \to NN\pi$
which can be written in the form of the Dalitz plot \cite{PDG02}~:
\begin{eqnarray}
       d \sigma_{N_1 N_2 \to N_3 N_4 \pi}^{med}
 &=& {(2m_N^*)^4 \over (2\pi)^3 64 q(\sqrt{s^*},m_N^*,m_N^*) (s^*)^{3/2}}
                                                      \nonumber \\
 &\times& \overline{|T_{N_1 N_2 \to N_3 N_4 \pi}|^2}(\sqrt{s}) S_{34}
   d(m_{34}^*)^2  d(m_{4\pi}^*)^2~,                   \label{sigmedswave}
\end{eqnarray}
where $(m_{34}^*)^2 = (p_3^* - p_4^*)^2$ 
and $(m_{4\pi}^*)^2 = (p_4^* - k)^2$ are the invariant masses squared 
of the pairs of outgoing particles $N_3,N_4$ and $N_4,\pi$ respectively.
$\overline{|T_{N_1 N_2 \to N_3 N_4 \pi}|^2}(\sqrt{s})$ is the vacuum
matrix element squared which depends on the vacuum c.m. energy $\sqrt{s}$
(see Eq.(\ref{srtst}) below) and on the isospins of the involved particles.
From Eq. (\ref{sigmedswave}) one gets the following
relation between the in-medium and vacuum cross sections~:
\begin{eqnarray}
       \sigma_{N_1 N_2 \to N_3 N_4 \pi}^{med}(\sqrt{s^*}) &=&
{ q(\sqrt{s},m_N,m_N) s^{3/2} \over q(\sqrt{s^*},m_N^*,m_N^*) (s^*)^{3/2} }
\left({ m_N^* \over m_N }\right)^4                         \nonumber \\
&\times& { \int\, d(m_{34}^*)^2 d(m_{4\pi}^*)^2
    \over \int\, d(m_{34})^2 d(m_{4\pi})^2 } 
\sigma_{N_1 N_2 \to N_3 N_4 \pi}^{vac}(\sqrt{s})~,       \label{sigren}
\end{eqnarray} 
where $\sigma_{N_1 N_2 \to N_3 N_4 \pi}^{vac}(\sqrt{s})$
is the vacuum cross section parameterized in Ref. \cite{Teis97}.

In order to maintain detailed balance, one must modify also the inverse 
processes $N\Delta \to NN$, $NR \to NN$, $\Delta\Delta \to NN$,
$\pi NN \to NN$ -- which are obviously essential for pion reabsorption --
simultaneously with the direct reactions.
We collect the formulas for the in-medium cross sections of the
inverse processes in Appendix B.

Since the mean field included in the BUU model is a Skyrme-like which
is different from the RMF model, a recipe must be found how to match the 
BUU kinematics of a nucleon-nucleon collision with the kinematics employing 
Dirac effective masses.  
One way is to take the incoming particle momenta from BUU and
construct $s^*$ by replacing the bare masses by the effective ones.
However, as we discussed in Sect. II, if the c.m. momentum is fixed,
a NN collision which is subthreshold for pion production in the vacuum may 
get above threshold in the nuclear medium. This would create practical 
difficulties in ensuring energy conservation for the outgoing particles.
Therefore, we have used the c.m. kinetic energy above the pion production
threshold from BUU as an input to the calculation of the
in-medium cross sections. Namely, for each NN collision we find
$Q = \sqrt{s} - 2m_N - m_\pi$, where $\sqrt{s}$ is calculated with
the incoming nucleon momenta from BUU using the vacuum dispersion
relation for nucleon energies. The in-medium c.m. energy $\sqrt{s^*}$ is
determined by requiring that $Q^* = Q$, i.e.
\begin{equation}
        \sqrt{s^*} = \sqrt{s} -2(m_N-m_N^*)~.            \label{srtst}
\end{equation}
Correspondingly, the c.m. momentum of the incoming particles gets
changed~: $q(\sqrt{s^*},m_N^*,m_N^*) \leq q(\sqrt{s},m_N,m_N)$.

Replacing only the vacuum cross sections by the in-medium ones
in BUU would, however, not correspond to a consistent description
of the resonance production/absorption rates because
\begin{equation}
   w_{12 \to 34} \propto v_{12}^* \sigma_{12 \to 34}^{med}  \label{trprob}
\end{equation}
where $w_{12 \to 34}$ is the transition probability per unit time.
\begin{equation}
   v_{12}^* = { \sqrt{s^*} \over \varepsilon_1^* \varepsilon_2^* }
              q(\sqrt{s^*},m_1^*,m_2^*)                   \label{v12}
\end{equation}
is the relative velocity and $\varepsilon_i^* = 
\sqrt{q^2 + (m_i^*)^2},~~i=1,2$ are the c.m. energies
of colliding particles 1 and 2.
Due to the difference between the mean fields in BUU and in the
RMF model, the relative velocities of 
colliding particles are, generally, different in both models too.
Therefore, in order to obtain the rates corresponding to the
RMF model, we have multiplied the cross sections calculated above
by the factor $v_{12}^*/v_{12}^{BUU}$, where $v_{12}^{BUU}$ is
the relative velocity of the colliding particles as given by BUU.

\begin{figure}

\vspace{1cm}

\includegraphics{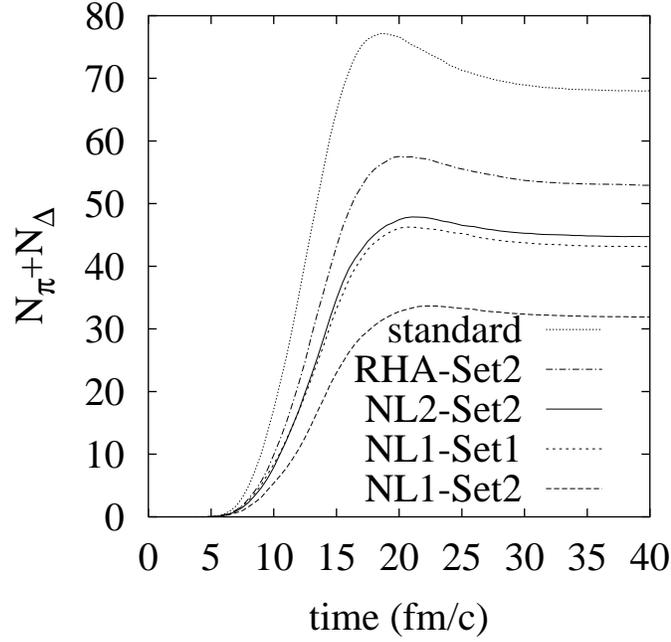}

\vspace{4cm}

\caption{\label{fig:pidelta} Time dependence of the sum of 
the pion and $\Delta$ multiplicities in the central collision
of Au+Au at 1.06 AGeV. The standard BUU calculation is shown by the 
dotted line. The dash-dotted, solid, short-dashed and long-dashed  
lines show the results with the in-medium modified resonance 
production/absorption cross sections by using the combinations
RHA-Set2, NL2-Set2, NL1-Set1 and NL1-Set2, respectively.}
\end{figure}

The BUU results have been obtained by applying the soft momentum
dependent mean field (SM, $K = 220$ MeV, see Refs. \cite{EffePhD,LCGM00}
for details) in the Hamiltonian propagation of baryons. 
Fig.~\ref{fig:pidelta} shows the time dependence of the sum 
of the pion and $\Delta$ multiplicities in a central collision of Au+Au at
1.06 A GeV. At $t=40$ fm/c practically all the $\Delta$'s have already
decayed, so that the final pion multiplicity can be determined from 
Fig.~\ref{fig:pidelta}. A standard calculation without any in-medium effects 
in the cross sections ends up with 68 pions, which is much more than
the experimental value of $41\pm7$ \cite{Pelte97}. Using the medium modified
cross sections reduces the pion multiplicity. A bigger reduction 
corresponds to a steeper dropping effective mass with the density 
(c.f. Fig.~\ref{fig:waltst}). Thus, the smallest pion multiplicity of 
$\sim 34$ is obtained within the NL1-Set2 combination. It is interesting that
the combinations NL1-Set1 and NL2-Set2 produce nearly equal pion numbers~: 
45 and 43, respectively. This is explained by the fact, that the maximum
central density reached in central Au+Au collisions at 1 A GeV
is about $2.4\rho_0$, while the cross sections given by the NL1-Set1
and NL2-Set2 combinations are similar at $\rho < 3\rho_0$ 
(see Fig.~\ref{fig:sigma_vs_ekcm_rho}). In all calculations below
we will always use the Set2 of the OPEM parameters, thus, we will drop
for brevity the label ``Set2''. 

\begin{figure}

\includegraphics{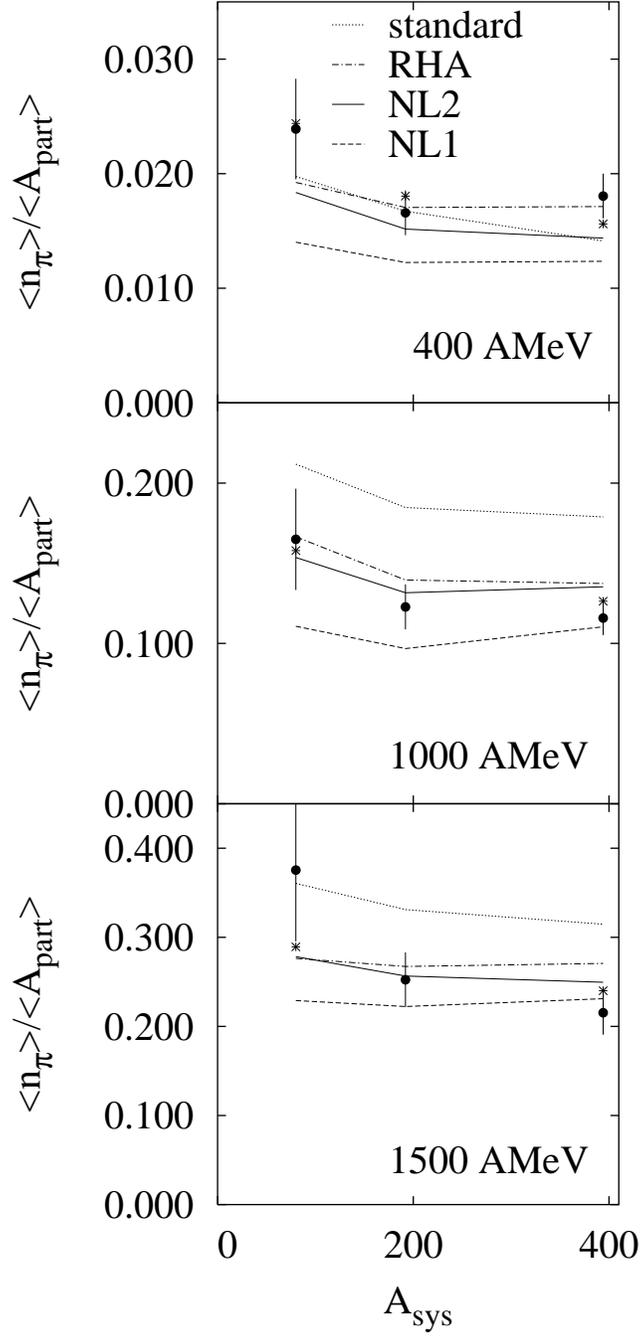}

\caption{\label{fig:pimul} Total pion yield per participating nucleon
for different energies and colliding systems. Dotted, dash-dotted,
solid and dashed lines correspond to the standard, RHA, NL2 and NL1
calculations, respectively. The data points are from Ref. \cite{Pelte97}.
The results of the data analysis A (B) are represented by filled circles
(stars) \cite{Pelte97}.}
\end{figure}

Fig.~\ref{fig:pimul} shows the average total pion multiplicity $<n_\pi>$ 
divided by the average participant number $<A_{part}>$ as a function
of the total mass number $A_{sys}$ of colliding nuclei in comparison 
to the data \cite{Pelte97}. The systems Ca+Ca, Ru+Ru and Au+Au were
studied at the beam energies of 400, 1000 and 1500 A MeV.
Following \cite{Pelte97}, we have fitted the total pion multiplicity as
a function of the participant number $A_{part}$ by a straight line
in the range $0.15 < A_{part}/A_{sys} < 0.85$. The participant number
$A_{part}$ has been extracted from the geometrical overlap of colliding 
nuclei for a given impact parameter taking into account smooth density 
profiles as described in \cite{Pelte97}.

First, we notice that there is no a single set of parameters which is able
to describe all data points in Fig.~\ref{fig:pimul} simultaneously.
The standard calculation agrees with the data at 400 A MeV, but it clearly
overpredicts the experiment at 1000 A MeV and, for heavy systems, also
at 1500 A MeV.

The NL1 parameterization works for the Au+Au system at 1000 and 1500 A MeV,
however, it is below the data at 400 A MeV. Furthermore, NL1 is 
unable to describe the experimental slopes at 1000 and 1500 A MeV. 

RHA is in the best agreement with data at 400 A MeV. Surprisingly, for
the Au+Au system at 400 A MeV RHA gives more pions than the standard
calculation. This, probably, can be explained by the reduced reabsorption
of $\Delta$'s in RHA due to the in-medium reduced cross section 
$\Delta N \to N N$. RHA is also rather close to the experiment at 1000 A MeV, 
while at 1500 A MeV the system mass dependence is too flat within RHA.

The best overall description of pion multiplicities seems to be reached
within the NL2 parameterization. More precise data, in particular, at 
the lowest energy are needed for definite conclusions, however. 

\begin{figure}

\includegraphics{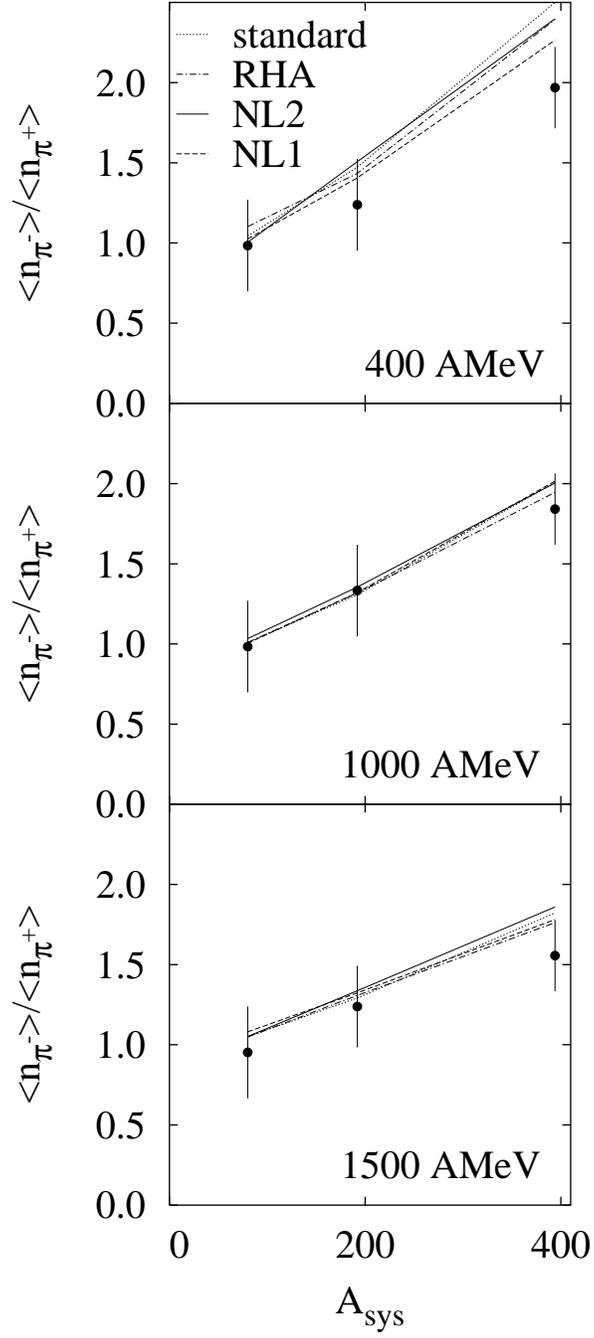}

\caption{\label{fig:pirat} Ratio of the charged pion yields for 
different energies and colliding systems. Various curves are denoted
as in Fig.~\ref{fig:pimul}. The data points are from Ref. \cite{Pelte97}.}

\end{figure}

Fig.~\ref{fig:pirat} shows the ratio of the average $\pi^-$ and $\pi^+$
multiplicities vs the total mass number of a system for Ca+Ca,
Ru+Ru and Au+Au at 400, 1000 and 1500 A MeV in comparison to the data
\cite{Pelte97}. The calculated ratio $<n_{\pi^-}>/<n_{\pi^+}>$ 
practically does not depend on the in-medium modifications of the cross 
sections, since these modifications do not change the ratios between the 
cross sections for various isospin channels. For Ca+Ca and Ru+Ru all the 
calculations agree with the data, while for the Au+Au system there is a 
systematic tendency to produce about 10 \% too many $\pi^-$'s in our 
calculations. Probably, the  $<n_{\pi^-}>/<n_{\pi^+}>$ ratio would be
described better if to take into account the isospin dependence of 
the in-medium part of the nucleon propagator (\ref{nucprop}).

\begin{figure}

\includegraphics{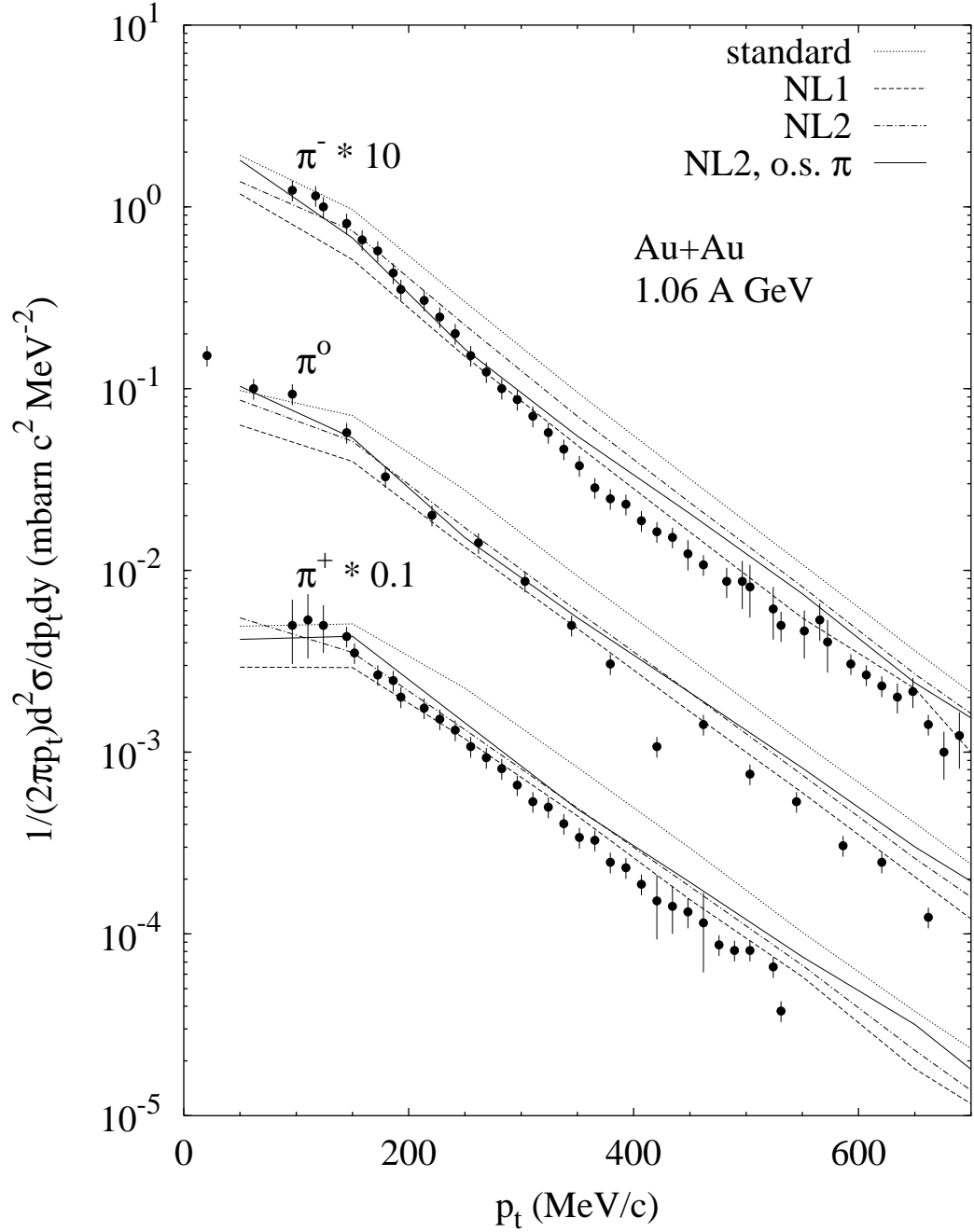}

\caption{\label{fig:piptsp_au106au} Pion inclusive transverse momentum 
spectra at midrapidity for the reaction Au+Au at 1.06 AGeV.
Dotted lines show the standard BUU calculation.
Dashed and dash-dotted lines show results with in-medium modified resonance 
production/absorption cross sections applying the NL1 and NL2 
parameterizations, respectively. Solid lines correspond to the
NL2 parameterization taking into account pion off-shellness.
The normalized rapidity intervals in which spectra are 
extracted are $Y^{(0)} = -0.2\div0.2$ for $\pi^\pm$ and 
$Y^{(0)} = -0.25\div0.21$ for $\pi^0$, where $Y^{(0)} \equiv 
(y/y_{proj})_{c.m.}$. The data are from Ref. \cite{Pelte97}.}
\end{figure}

Fig.~\ref{fig:piptsp_au106au} shows the inclusive transverse momentum spectra 
of pions at midrapidity for the Au+Au system at 1.06 A GeV.  
The standard calculation (dotted line) overpredicts 
the data at $p_t > 150$ MeV/c. NL1 (dashed line) leads to a good description 
at $p_t > 200$ MeV/c, but it underpredicts the experiment at low 
$p_t$. Almost the same result has been obtained earlier in \cite{LCLM01}
using the phenomenological density dependent quenching factor for
the $NN \leftrightarrow N\Delta$ cross sections. A better overall description
is reached within the NL2 parameterization (dash-dotted line), however the 
yields of $\pi^-$'s and $\pi^0$'s are still underpredicted in the region of 
small transverse momenta. This is improved by taking into account the pion 
off-shellness in BUU \cite{LM02} (solid line), which enhances the soft pion 
yield keeping the total pion number practically unchanged. 

\begin{figure}

\vspace{1cm}

\includegraphics{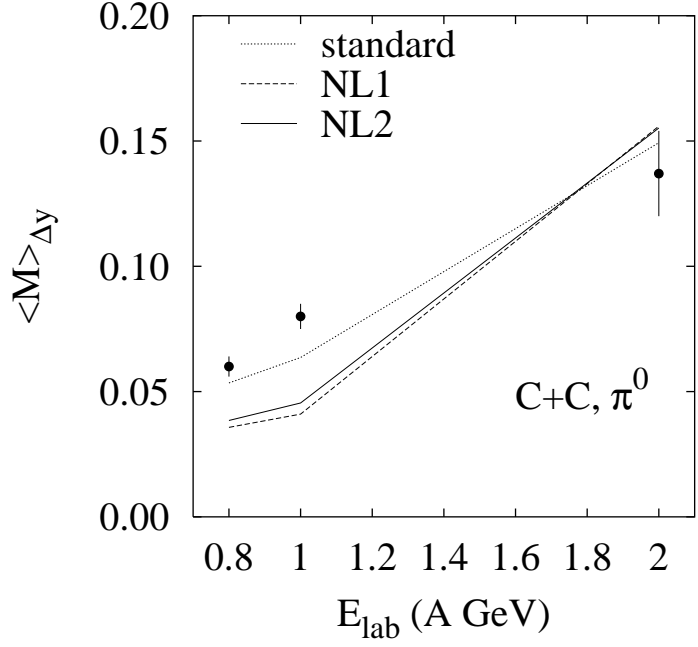}

\vspace{4cm}

\caption{\label{fig:pinum_cc} The average $\pi^0$ multiplicity in 
a narrow rapidity interval near midrapidity as a function of the beam 
energy for C+C collisions. Dotted, dashed and solid lines correspond to the 
standard, NL1 and NL2 calculations, respectively.
The rapidity intervals are $y=0.42\div0.74$ at $E_{lab}=0.8$ and $1.0$ A GeV,
and  $y=0.80\div1.08$ at $E_{lab}=2$ A GeV. 
The data are from Ref. \cite{Aver97}.}
\end{figure}

In order to see how the density modified cross sections work in the case
of light colliding nuclei we selected the TAPS data on $\pi^0$ production
from collisions of C+C at 0.8, 1.0 and 2.0 A GeV \cite{Aver97}. 
Fig.~\ref{fig:pinum_cc} shows the impact parameter averaged $\pi^0$
multiplicity $<M>_{\Delta y}$ in a narrow rapidity interval near 
$y=0$. Standard calculation is in the best agreement with the data for 
$<M>_{\Delta y}$, while the NL1 and NL2 calculations are
below the data for the beam energies of 0.8 and 1 A GeV.

\begin{figure}

\includegraphics{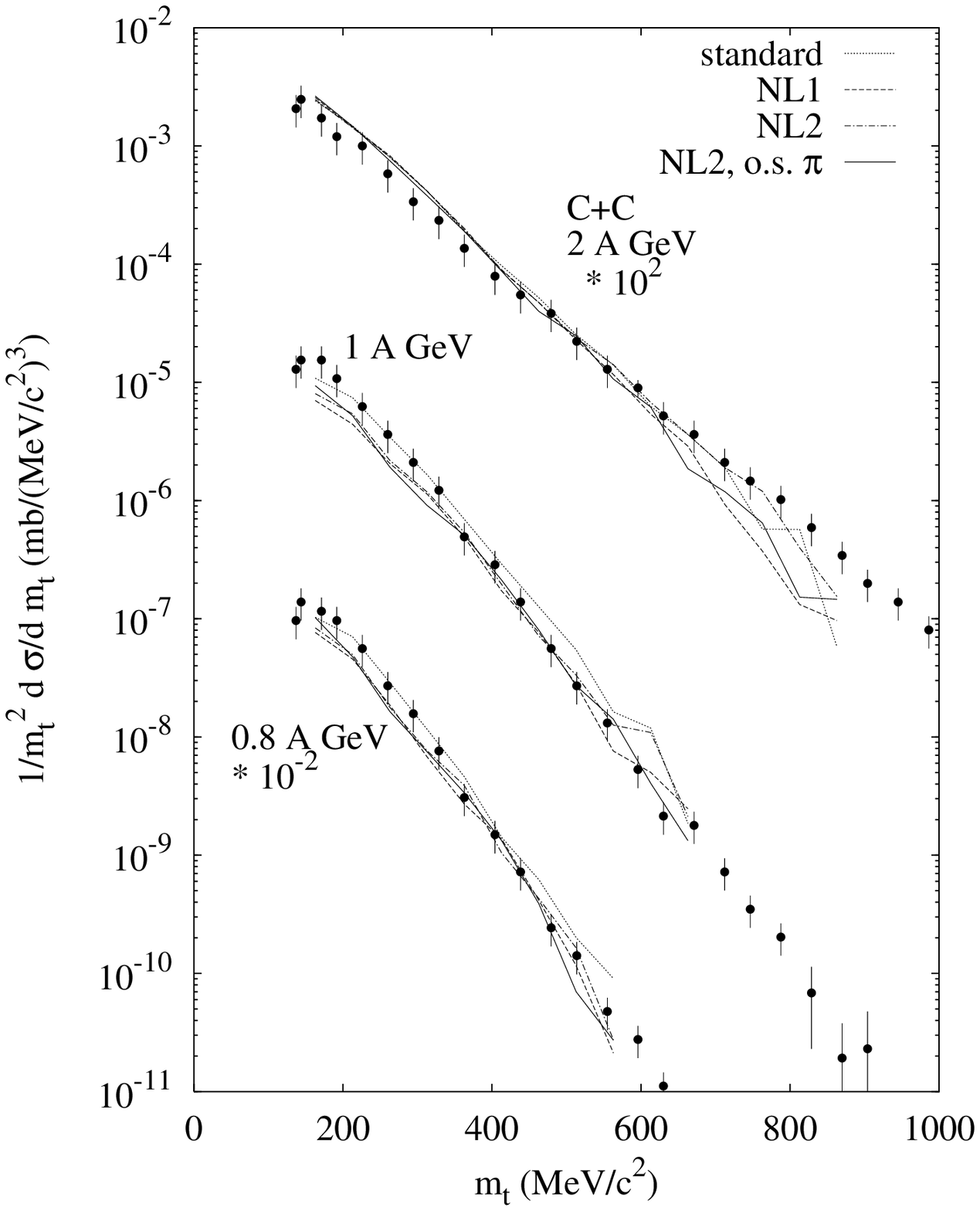}

\caption{\label{fig:dmtpi_cc_new} Transverse mass spectra of $\pi^0$ mesons
in the narrow rapidity intervals near midrapidity 
(see caption to Fig.~\ref{fig:pinum_cc}) for the $^{12}$C+$^{12}$C system at 
0.8, 1.0 and 2.0 A GeV. Curves are denoted as in 
Fig.~\ref{fig:piptsp_au106au}. The data are from Ref. \cite{Aver97}.}
\end{figure}

Finally, Fig.~\ref{fig:dmtpi_cc_new} shows transverse mass spectra
of $\pi^0$'s at midrapidity for the C+C collisions at 0.8, 1 and 2 A GeV.
The standard calculation provides a quite good description of the
data \cite{Aver97}. In-medium modifications of the cross sections 
do not destroy a good description at 2 A GeV and also improve agreement
with experiment for large transverse masses for 0.8 and 1 A GeV,
but tend to underpredict the data at small $m_t$ for 0.8 and 1 A GeV.

\section{ Summary }

In this work we have calculated the in-medium $NN \to N\Delta(1232)$
cross section within the OPEM taking into account the exchange pion
collectivity, vertex corrections by the contact nuclear interactions
and the effective masses of the nucleon and $\Delta$ resonance.

(i) We have observed that even without the effective mass modifications
the cross section decreases with the nuclear matter density at
high densities if one takes into account the in-medium $\Delta$ width and 
includes the $NN^{-1}$ Lindhard function in the calculations 
(Fig.~\ref{fig:sigma_vs_rho}).

(ii) Inclusion of the effective mass modifications for the nucleons and 
$\Delta$'s leads to an additional strong reduction of the cross section. 

(iii) The FOPI data \cite{Pelte97} on the total pion multiplicity 
from the systems Ca+Ca, Ru+Ru and Au+Au at 0.400, 1.000 and 1.500 A GeV
seem to require a dropping effective mass with the baryon density 
(NL2 version of the RMF model) in combination with the universal value 
of 0.6 for all the Landau-Migdal parameters in the spin-isospin channel.
We note that this still depends somewhat on the size of the 
$\Delta$-spreading width used (80 MeV $\rho/\rho_0$ \cite{Hir79,OS87}).
More precise data on the total pion multiplicity are needed in order to 
better select the model for the density dependence of the Dirac effective 
masses of the baryons.

The effect of the medium modifications of the $NN \leftrightarrow N\Delta$ 
cross sections on the pion multiplicity depends also on the assumption about 
other channels of the pion production/absorption in NN collisions, most 
importantly, on the $s$-wave direct channel $NN \leftrightarrow NN\pi$.
Including the effective mass modifications in the $NN \leftrightarrow 
N\Delta$ channel {\it only}, does not lead to strong enough reduction
of the observed pion multiplicity, since then more pions are produced
in the $s$-wave channel.
The in-medium modifications of the higher resonance cross sections
do not influence the pion production at 1-2 A GeV collision energy
sensitively~: other particles like $\eta$ and $\rho$ mesons are, 
probably, more sensitive to higher resonance in-medium modifications.
 
The calculations with the in-medium modified cross sections slightly 
underpredict the low transverse mass $\pi^0$ yield at midrapidity for 
C+C collisions at 0.8 and 1.0 A GeV (Fig.~\ref{fig:dmtpi_cc_new}). 
This implies, that with decreasing mass number of the colliding system 
the in-medium corrections disappear faster than in our model. 
The reason could be a strong nonequilibrium momentum distribution
at the beginning of collision: instead of one Fermi sphere
there are two nonoverlapping Fermi spheres, which is not taken into 
account in our calculation of the effective mass. Indeed, the
nonequilibrium momentum distribution reduces the scalar density,
and, therefore, increases the effective mass. This effect is expected
to be stronger for lighter systems, since smaller baryon density
is reached in this case. 

Another open problem is the general tendency of the transport models to 
overpredict pion multiplicity in heavy colliding systems at AGS energies 
(c.f. Ref. \cite{Weber02}).
This problem can not be solved by just density modifications of the resonance 
production/absorption cross sections, since most pions are produced by the 
string mechanism here. In-medium modifications of the FRITIOF model, thus, 
could help to solve this problem.

\begin{acknowledgments}
Numerous stimulating discussions with M. Post and Dr. L. Alvarez-Ruso 
are gratefully acknowledged. The authors are grateful to Markus Post
for the reading of the manuscript before publication, critical
comments and suggestions and to Luis Alvarez-Ruso for pointing out
Ref. \cite{Fernandez95} which has influenced our results. 
\end{acknowledgments}

\appendix

\section{ Calculation of $\overline{|T^{med}|^2}$ for
          the process $pp \to n\Delta^{++}$ }

For the calculation of $\overline{|T^{med}|^2}$ it is convenient to 
introduce the renormalized coupling constants
\begin{eqnarray}
\tilde{f}_\Delta(k) &\equiv& {\det}^{-1}(\hat A(k)) 
    [1 + (g_{NN}^\prime-g_{N\Delta}^\prime)\chi_N(k)] 
f_\Delta(k^2) ~,                                             \label{fdren} \\
\tilde{f}(k) &\equiv& {\det}^{-1}(\hat A(k)) 
    [1 + (g_{\Delta\Delta}^\prime-g_{N\Delta}^\prime)\chi_\Delta(k)] 
f(k^2)                                                       \label{fren}
\end{eqnarray}
and the renormalized Landau-Migdal parameter
\begin{equation}
\tilde{g}(k) \equiv { \det(\hat A(k)) g_{N\Delta}^\prime \over
[1 + (g_{NN}^\prime-g_{N\Delta}^\prime)\chi_N(k)] 
[1 + (g_{\Delta\Delta}^\prime-g_{N\Delta}^\prime)\chi_\Delta(k)]}~. 
                                                           \label{gren}
\end{equation}
In terms of $\tilde{f}_\Delta(k)$, $\tilde{f}(k)$ and $\tilde{g}(k)$
the renormalized vertex operators and the renormalized contact interaction
have the same form as the vacuum ones~:
\begin{eqnarray}
        \tilde\Gamma_j^{abs}(k) &=& -{\tilde{f}_\Delta(k) \over m_\pi}
        ({\bf S \cdot k}) T_j~,                       \label{gama1}   \\
        \tilde\Gamma_j^{dec}(k) &=& {\tilde{f}(k) \over m_\pi}
(\mbox{\boldmath${\mathbf\sigma}$\unboldmath}{\bf \cdot k}) \tau_j~,
                                                      \label{gamd1} \\
\tilde{V}_{4,3;2,1}(k) &=& {\tilde{f}_\Delta(k)\tilde{f}(k) \over m_\pi^2}
 \tilde{g}(k)
        [\chi^\dag_\Delta(4) S_\alpha {\bf T} \chi(2)]
        [\chi^\dag(3) \sigma_\alpha 
         \mbox{\boldmath${\mathbf\tau}$\unboldmath} \chi(1)]~.
                                                               \label{Vcont1}
\end{eqnarray}
Notice, that $\tilde{f}_\Delta(k)$, $\tilde{f}(k)$ and $\tilde{g}(k)$
are complex-valued quantities. Substituting (\ref{gama1}),(\ref{gamd1})
and (\ref{Vcont1}) into (\ref{Ta}) on the place of corresponding vacuum
quantities and also replacing  $({\bf S \cdot k}) \to ({\bf S \cdot k_4})$ 
(see Eq.(\ref{k4})) in Eq.(\ref{gama1}) one can get
after some algebra~:
\begin{eqnarray}
&\overline{|T_a^{med}|^2}& =         
   {1 \over 4} \sum_{\lambda_1,\lambda_2,\lambda_3,\lambda_{\Delta4}}\,
   |T_a^{med}|^2  =  {4 \over 3} 
   {|\tilde{f}_\Delta(k)|^2 |\tilde{f}(k)|^2 \over m_\pi^4}    \nonumber \\
&\times& \left[ {\bf k_4}^2{\bf k}^2 |D(k)|^2                     
         + ( \tilde{g}^*(k)D(k) + \tilde{g}(k)D^*(k) ) ({\bf k_4 \cdot k})
                   + 3 |\tilde{g}(k)|^2 \right]
         \left(-{t \over {\bf k}^2}\right)~,        \label{Tamedsq}
\end{eqnarray}
where the term $-t/{\bf k}^2$ is introduced in order to restore the
Lorentz invariance, as discussed in Sect. II. The exchange term
$\overline{|T_b^{med}|^2}$ is given by Eq.(\ref{Tamedsq}) with
replacements $k \to k^\prime$, $t \to u$ and 
${\bf k_4} \to {\bf k_4^\prime}$ (see Eq.(\ref{k4p})). 

For the spin-averaged interference term we get~:
\begin{eqnarray}
 &{1 \over 4}& \sum_{\lambda_1,\lambda_2,\lambda_3,\lambda_{\Delta4}}\,
[T_a^{med} (T_b^{med})^* + (T_a^{med})^* T_b^{med}]       
 = - {\tilde{f}_\Delta^*(k^\prime)\tilde{f}^*(k^\prime)
         \tilde{f}_\Delta(k)\tilde{f}(k) \over m_\pi^4}  \nonumber \\
 &\times& \left\{ {1 \over 3}[ 
2({\bf k_4^\prime \cdot k_4})({\bf k^\prime \cdot k}) + 
({\bf k_4 \cdot k}) ({\bf k_4^\prime \cdot k^\prime}) - 
({\bf k_4 \cdot k^\prime})({\bf k_4^\prime \cdot k}) ] 
  D^*(k^\prime)  D(k)  \right.                   \nonumber \\
 &+& \left. {4 \over 3} ({\bf k_4^\prime \cdot k^\prime}) D^*(k^\prime) 
                        \tilde{g}(k)
  +  {4 \over 3} ({\bf k_4 \cdot k}) D(k) \tilde{g}^*(k^\prime)     
  +  4\tilde{g}^*(k^\prime)\tilde{g}(k) \right\} 
{(tu)^{1/2} \over ({\bf k}^2{\bf k^\prime}^2)^{1/2}}  
        + c.c.~.                                       \label{Tintmedsq}
\end{eqnarray}

\section{ In-medium cross sections for the inverse processes }

We use below the detailed balance principle, i.e. the equality of the matrix
elements of the direct and inverse processes, in order to get the
cross sections for the reactions $N\Delta \to NN$, $NR \to NN$, 
$\Delta\Delta \to NN$ and the pion  $s$-wave absorption rate 
by two-nucleon pairs $\pi NN \to NN$.

The in-medium differential cross section for the process $N\Delta \to NN$ 
is given by the expression (c.f. Eq.(\ref{sigmedcm}))~:
\begin{equation} 
   {d\sigma_{N_3\Delta_4 \to N_1N_2}^{med}(\sqrt{s^*}) \over d\Omega}     
= { (2m_N^*)^3 2M_\Delta^* \over 64 \pi^2  s^*} {2 \over 2J_\Delta + 1}
  \overline{|T^{med}|^2}
{ q(\sqrt{s^*},m_N^*,m_N^*) \over q(\sqrt{s^*},M_\Delta^*,m_N^*)}
S_{12}~,                                                \label{DNNN}
\end{equation}
where $\overline{|T^{med}|^2}$ is calculated in Appendix A 
(Eqs. (\ref{Tamedsq}),(\ref{Tintmedsq})). $J_\Delta = 3/2$ is the spin
of the $\Delta$-resonance. For the process $NR \to NN$, where $R$ means 
one of the higher resonances, one should simply replace in Eq.(\ref{DNNN}) 
``$\Delta$'' by ``$R$''~: now $M_R^*$ and $J_R$ are the effective 
mass and spin of the resonance $R$, respectively. $\overline{|T^{med}|^2}$ 
is now a constant depending only on the isospin channel.

The in-medium cross section of the reaction $\Delta\Delta \to NN$ reads 
as follows (c.f. Eq.(\ref{sigmedDDcm})~:
\begin{eqnarray}
   \sigma_{\Delta_3\Delta_4 \to N_1N_2}^{med}(\sqrt{s^*})
&=& {(2m_N^*)^2 (2m_\Delta^*)^2 \over 16\pi s^*} {4 \over (2J_\Delta +1)^2} 
  \overline{|T_{N_1N_2 \to \Delta_3\Delta_4}|^2}         \nonumber \\
&\times& { q(\sqrt{s^*},m_N^*,m_N^*) \over q(\sqrt{s^*},M_3^*,M_4^*)} 
S_{12}~.                                                   \label{DDNN}
\end{eqnarray}

The pion $s$-wave absorption rate by the two-nucleon pairs is~:
\begin{eqnarray}
   & &\Sigma_\pi^{med~>}(k) = {1 \over 2k_0} \sum_{1,2}
\int\, {g d^3p_1 \over (2\pi)^3 2 \varepsilon_1^*} n_1
       {g d^3p_2 \over (2\pi)^3 2 \varepsilon_2^*} n_2      
       {g d^3p_3 \over (2\pi)^3 2 \varepsilon_3^*} (1-n_3)
           {g d^3p_4 \over (2\pi)^3 2 \varepsilon_4^*} (1-n_4) \nonumber \\
 &\times&  (2\pi)^4 \delta^{(4)}(k + p_1^* + p_2^* - p_3^* - p_4^*)
           S_{12} S_{34} (2m_N^*)^4  
           \overline{|T_{N_3 N_4 \to N_1 N_2 \pi}|^2}(\sqrt{s})~,
                                                        \label{swpimed}
\end{eqnarray}
where $g=2$ is the nucleon spin degeneracy factor, 
$n_i \equiv n({\bf p}_i),~~(i=1,2,3,4)$ are the nucleon occupation numbers.
The sum in Eq.(\ref{swpimed}) is taken over
the isospin projections of the incoming nucleons 1 and 2.
The vacuum matrix element squared
\begin{equation}
   \overline{|T_{N_3 N_4 \to N_1 N_2 \pi}|^2}(\sqrt{s})
= { (2\pi)^3 64 q(\sqrt{s},m_N,m_N) s^{3/2} 
    \sigma_{N_3 N_4 \to N_1 N_2 \pi}^{vac}(\sqrt{s}) 
\over (2m_N)^4 S_{12} \int\, d(m_{12})^2 d(m_{2\pi})^2 }   \label{swmelem}
\end{equation}
depends on the vacuum c.m. energy and on the particle isospins. The integral 
(\ref{swpimed}) is performed by the Monte-Carlo
method as described in Ref. \cite{EffePhD}.

\newpage

\begin{description}
\item[Table 1.] Parameters of the medium modified OPEM
\end{description}

\vspace{0.5cm}

\begin{center}
\begin{tabular}{|c|c|c|c|c|}
\hline
Set \# & $g_{NN}^\prime$ & $g_{N\Delta}^\prime$ & $g_{\Delta\Delta}^\prime$ &
$\Lambda$, GeV \\
\hline
1  & 0.9 & 0.4 & 0.4 & 0.70 \\
2  & 0.6 & 0.6 & 0.6 & 0.67 \\
\hline
\end{tabular}
\end{center}

\clearpage

\thispagestyle{empty}

\clearpage

\thispagestyle{empty}

\begin{figure}

\includegraphics{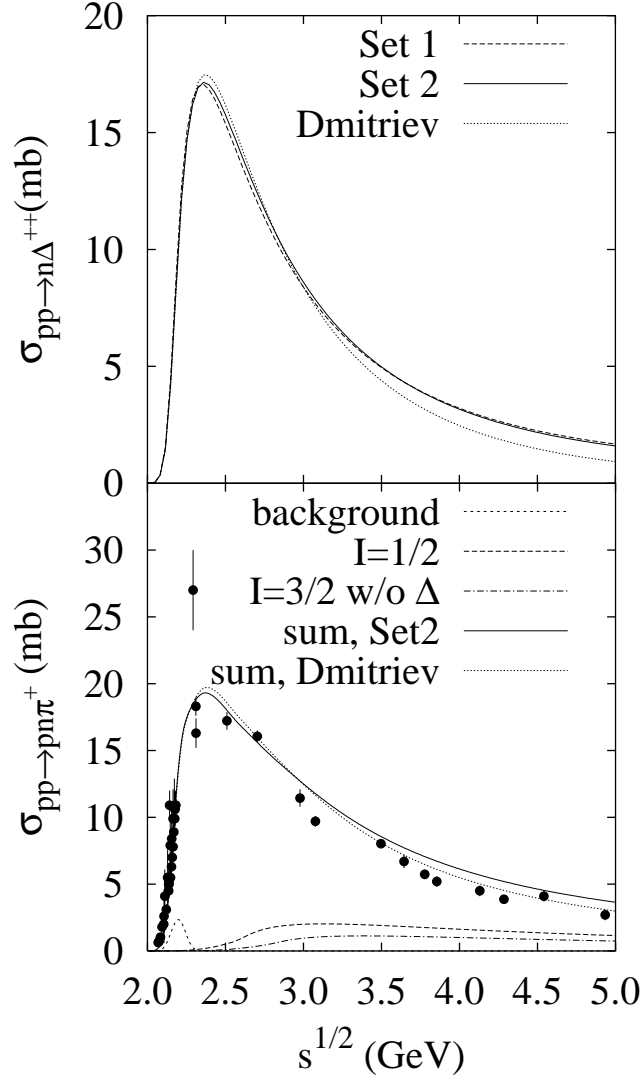}

\caption{\label{fig:sigma_vs_plab} Upper panel: the total cross section
$pp \to n\Delta^{++}$ in vacuum as a function of the c.m. energy
$\sqrt{s}$.
Calculations are performed using the Set 1 (dashed line) and Set 2 
(solid line) of the OPEM parameters (see Table 1). Results 
obtained within the model of Ref. \cite{Dm86} are shown by dotted line.
Lower panel: vacuum cross section of a single $\pi^+$ production in $pp$ 
collisions vs $\sqrt{s}$. Results with the $pp \to n\Delta^{++}$ cross 
section calculated with the Set 2 and using the model of Ref. \cite{Dm86} 
are shown by solid and dotted lines respectively. The $s$-wave background, 
isospin 1/2 and 3/2 (excluding $P_{33}(1232)$) resonance contributions 
\cite{Teis97} are shown by short-dashed, long-dashed and dash-dotted lines 
respectively. Solid and dotted lines show the incoherent sum of the 
$pp \to n\Delta^{++}$ cross section multiplied by 10/9 and the background 
plus isospin 1/2 and 3/2 higher resonance contributions.
The data are from Ref. \cite{Bald87}.}
\end{figure}

\clearpage

\thispagestyle{empty}

\begin{figure}

\vspace{5cm}

\includegraphics{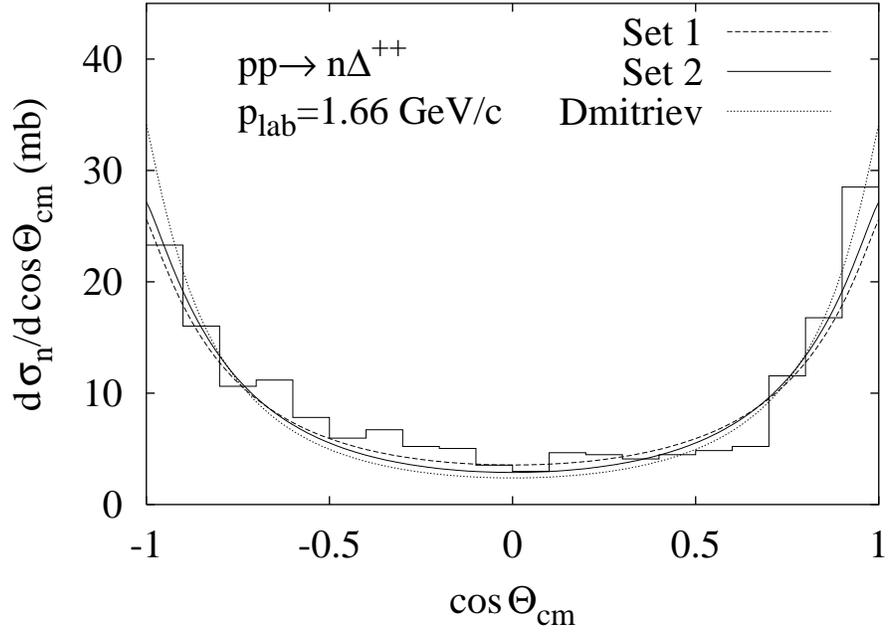}

\vspace{7cm}

\caption{\label{fig:sigma_vs_costh} Outgoing neutron c.m. polar angle 
distribution in the reaction $pp \to np\pi^+$ at $p_{lab}=1.66$ GeV/c. 
In the calculations only the $pp \to n\Delta^{++}$ channel is taken
into account. 
The curves are labeled as in the upper panel of Fig.~\ref{fig:sigma_vs_plab}. 
The histogram shows the data from Ref.~\cite{Bugg64}.}
\end{figure}

\clearpage

\thispagestyle{empty}

\clearpage

\thispagestyle{empty}

\clearpage

\thispagestyle{empty}

\clearpage

\thispagestyle{empty}

\clearpage

\thispagestyle{empty}

\clearpage

\thispagestyle{empty}

\clearpage

\thispagestyle{empty}

\begin{figure}

\vspace{-1cm}

\includegraphics{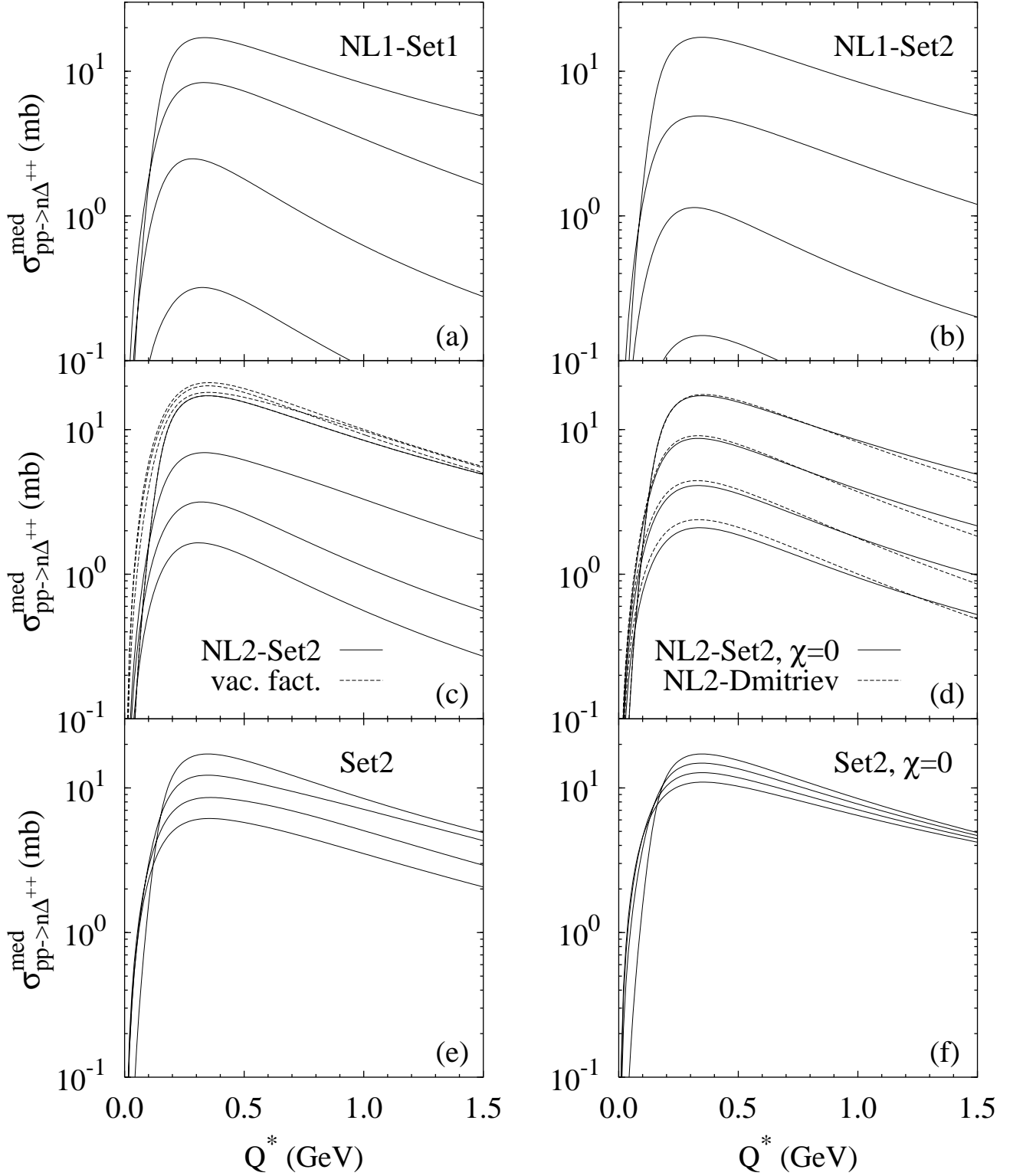}


\caption{\label{fig:sigma_vs_ekcm_rho} In-medium $pp \to n\Delta^{++}$
cross section as a function of the kinetic energy above the pion production
threshold $Q^*$ (see Eq.(\ref{Qstar})) for various parameter sets
(see text for details). In all cases, excepting the dashed lines
in panel (c), lines from the uppermost
to the lowermost correspond to the baryon densities 
$\rho = 0,~\rho_0,~2\rho_0$ and $3\rho_0$, respectively. For the dashed lines
in panel (c) the density order is opposite.}
\end{figure}

\clearpage

\thispagestyle{empty}

\clearpage

\thispagestyle{empty}

\clearpage

\thispagestyle{empty}

\clearpage

\thispagestyle{empty}

\clearpage

\thispagestyle{empty}

\clearpage

\thispagestyle{empty}

\clearpage

\thispagestyle{empty}

\clearpage

\thispagestyle{empty}

\newpage
    
\begin{thebibliography}{99}

\bibitem{BDG88} G.F. Bertsch and S. Das Gupta, 
                Phys. Rep. {\bf 160}, 189 (1988).

\bibitem{Cas90} W. Cassing, V. Metag, U. Mosel, and K. Niita,
                Phys. Rep. {\bf 188}, 363 (1990). 

\bibitem{Mosel91} U. Mosel, Annu. Rev. Nucl. Part. Sci., 
                  {\bf 41}, 29 (1991).

\bibitem{Ehe93} W. Ehehalt, W. Cassing, A. Engel, U. Mosel,
                and Gy. Wolf, Phys. Rev. C {\bf 47}, R2467 (1993).

\bibitem{Bass95} S.A. Bass, C. Hartnack, H. Stoecker, W. Greiner,
                 Phys. Rev. C {\bf 51}, 3343 (1995).

\bibitem{Ko96} C.M. Ko and G.Q. Li, J. Phys. {\bf G 22}, 1673 (1996).

\bibitem{Teis97} S. Teis, W. Cassing, M. Effenberger, A. Hombach,
U. Mosel and Gy. Wolf, Z. Phys. A {\bf 356}, 421 (1997);\\
S. Teis, W. Cassing, M. Effenberger, A. Hombach,
U. Mosel and Gy. Wolf, Z. Phys. A {\bf 359}, 297 (1997).

\bibitem{HR97} J. Helgesson and J. Randrup, 
               Phys. Lett. B {\bf 411}, 1 (1997).

\bibitem{Uma98} V.S. Uma Maheswari, C. Fuchs, Amand Faessler,
L. Sehn, D.S. Kosov, Z. Wang, Nucl. Phys. A {\bf 628}, 669 (1998).

\bibitem{CB99} W. Cassing and E.L. Bratkovskaya, 
               Phys. Rep. {\bf 308}, 65 (1999).

\bibitem{Weber02} H. Weber, E.L. Bratkovskaya, W. Cassing, H. St\"ocker,
                  Phys. Rev. C {\bf 67}, 014904 (2003).

\bibitem{Pelte97} D. Pelte et al.,
                  Z. Phys. A {\bf 357}, 215 (1997); 
                  M.R. Stockmeier et al., 
                  GSI Scientific Report 2001, p. 35.

\bibitem{LCLM01} A.B. Larionov, W. Cassing, S. Leupold, U. Mosel,
                 Nucl. Phys. A {\bf 696}, 747 (2001).

\bibitem{tHM87} B. ter Haar and R. Malfliet,
                Phys. Rev. C {\bf 36}, 1611 (1987).

\bibitem{Bertsch88} G.F. Bertsch, G.E. Brown, V. Koch and B.-A. Li,
                    Nucl. Phys. A {\bf 490}, 745 (1988).

\bibitem{WuKo89} J.Q. Wu and C.M. Ko, 
                 Nucl. Phys. A {\bf 499}, 810 (1989).

\bibitem{HR95} J. Helgesson and J. Randrup, 
               Ann. Phys. (N.Y.) {\bf 244}, 12 (1995).

\bibitem{EW88} T. Ericson and W. Weise, {\cal Pions and Nuclei},
               Clarendon Press, Oxford, 1988.

\bibitem{Dm86} V. Dmitriev, O. Sushkov and C. Gaarde,
               Nucl. Phys. A {\bf 459}, 503 (1986).

\bibitem{BW75} G.E. Brown and W. Weise,
               Phys. Rep. {\bf 22}, 279 (1975).

\bibitem{MeyerTV81} J. Meyer-Ter-Vehn, 
                    Phys. Rep. {\bf 74}, 323 (1981).

\bibitem{Mig90} A.B. Migdal, E.E. Saperstein, M.A. Troitsky
                and D.N. Voskresensky, 
                Phys. Rep. {\bf 192}, 179 (1990).

\bibitem{Oset79} E. Oset and M. Rho, 
                 Phys. Rev. Lett. {\bf 42}, 47 (1979).

\bibitem{Koerfgen97} B. K\"orfgen, P. Oltmanns, F. Osterfeld
                     and T. Udagawa, 
                     Phys. Rev. C {\bf 55}, 1819 (1997).

\bibitem{Arve94} P. Arve and J. Helgesson, 
                 Nucl. Phys. A {\bf 572}, 600 (1994).

\bibitem{BD} J.D. Bjorken and S.D. Drell, {\it Relativistic quantum
             mechanics}, McGraw-Hill, New York, 1965.  

\bibitem{EBM99} M. Effenberger, E.L. Bratkovskaya, and U. Mosel, 
                Phys. Rev. C {\bf 60}, 44614 (1999).

\bibitem{Fernandez95} P. Fern\'andez de C\'ordoba, E. Oset, 
M.J. Vicente-Vacas, Yu. Ratis, J. Nieves, B. L\'opez-Alvaredo, F. Gareev,
                Nucl. Phys. A {\bf 586}, 586 (1995).

\bibitem{Bald87} A. Baldini et al., {\cal Landolt-B\"ornstein}, V. 12,
                 Springer Verlag, Berlin, 1987.

\bibitem{Bugg64} D.V. Bugg et al., Phys. Rev. {\bf 133}, B1017 (1964).

\bibitem{Wehr89} K. Wehrberger, C. Bedau and F. Beck,
                 Nucl. Phys. A {\bf 504}, 797 (1989).

\bibitem{Lee86} S.J. Lee et al., Phys. Rev. Lett. {\bf 57}, 2916 (1986).

\bibitem{SW86} B.D. Serot and J.D. Walecka, Adv. Nucl. Phys. {\bf 16},
               1 (1986).

\bibitem{Hir79} M. Hirata, J.H. Koch, F. Lenz and E.J. Moniz,
                Ann. Phys. (N.Y.) {\bf 120}, 205 (1979).

\bibitem{OS87} E. Oset and L.L. Salcedo,
               Nucl. Phys. A {\bf 468}, 631 (1987).

\bibitem{RW94} R. Rapp and J. Wambach, Nucl. Phys. A {\bf 573}, 626 (1994).

\bibitem{Kim97} H. Kim, S. Schramm, S.H. Lee, 
                Phys. Rev. C {\bf 56}, 1582 (1997).

\bibitem{FW71} A.L. Fetter and J.D. Walecka, {\it Quantum theory
               of many-particle systems}, McGraw-Hill, New York, 1971.

\bibitem{OP81} E. Oset and A. Palanques-Mestre,
               Nucl. Phys. A {\bf 359}, 289 (1981).

\bibitem{Oset90} E. Oset, P. Fern\'andez de C\'ordoba, L.L. Salcedo
                 and R. Brockmann,
                 Phys. Rep. {\bf 188}, 79 (1990).

\bibitem{Xia88} L.H. Xia et al., Nucl. Phys. A {\bf 485}, 721 (1988).

\bibitem{Li93} G.Q. Li and R. Machleidt, 
               Phys. Rev. C {\bf 48}, 1702 (1993);\\
               G.Q. Li and R. Machleidt, 
               Phys. Rev. C {\bf 49}, 566 (1994).

\bibitem{Fuchs01} C. Fuchs, A. Faessler, and M. El-Shabshiri,
                   Phys. Rev. C {\bf 64}, 024003 (2001).

\bibitem{EffePhD}  M. Effenberger, PhD thesis, Uni. Giessen, 1999,\\
http://theorie.physik.uni-giessen.de/html/dissertations.html.

\bibitem{PDG02} K. Hagiwara et al., Phys. Rev. D {\bf 66}, 010001 (2002). 

\bibitem{LCGM00} A.B. Larionov, W. Cassing, C. Greiner, and U. Mosel,
                 Phys. Rev. C {\bf 62}, 064611 (2000).

\bibitem{LM02} A.B. Larionov and U. Mosel, 
               Phys. Rev. C {\bf 66}, 034902 (2002).

\bibitem{Aver97} R. Averbeck et al., Z. Phys. A {\bf 359}, 65 (1997).


\end{thebibliography}
\end{document}